\documentclass[12pt,preprint]{aastex}

\begin{document}

\title{Finding the Instability Strip for 
Accreting Pulsating White Dwarfs from HST and Optical Observations
\footnote{Based on observations made with the 
NASA/ESA Hubble Space
Telescope, obtained at the Space Telescope Science Institute, which is
operated by the Association of Universities for Research in Astronomy, Inc.,
under NASA contract NAS 5-26555, and with the Apache Point Observatory 3.5m 
telescope which is owned and operated by the Astrophysical Research
Consortium.}}

\author{Paula Szkody\altaffilmark{2},
Anjum Mukadam\altaffilmark{2},
Boris T. G\"ansicke\altaffilmark{3},
Arne Henden\altaffilmark{4},
Matthew Templeton\altaffilmark{4},
Jon Holtzman\altaffilmark{5},
Michael H. Montgomery\altaffilmark{6},
Steve B. Howell\altaffilmark{7},
Atsuko Nitta\altaffilmark{8},
Edward M. Sion\altaffilmark{9},
Richard D. Schwartz\altaffilmark{10},
William Dillon\altaffilmark{4}}

\altaffiltext{2}{Department of Astronomy, University of Washington, Box 351580,
Seattle, WA 98195}
\altaffiltext{3}{Dept. of Physics, University of Warwick, Coventry CV4 7AL, UK}
\altaffiltext{4}{American Association of Variable Star Observers, 25 Birch Street, Cambridge, MA 02138}
\altaffiltext{5}{Department of Astronomy, New Mexico State University, Box
30001, Las Cruces, NM 88003}
\altaffiltext{6}{Department of Astronomy, University of Texas, C1400, Austin, TX 78712}
\altaffiltext{7}{NOAO, 950 N. Cherry Ave., Tucson, AZ 85719}
\altaffiltext{8}{Gemini Observatory, 670 N A'ohoku Pl., Hilo, HI 96720 and
Subaru Telescope, 650 N A'ohoku Pl., Hilo, HI 96720}
\altaffiltext{9}{Dept. of Astronomy \& Astrophysics, Villanova University, 
Villanova, PA 19085}
\altaffiltext{10}{Galaxy View Observatory, Sequim, WA 98382}

\begin{abstract}
Time-resolved low resolution Hubble Space Telescope ultraviolet spectra together with 
ground-based optical photometry and spectra are used to constrain the
temperatures and pulsation properties of six
cataclysmic variables containing pulsating white dwarfs. Combining our
temperature determinations for the five pulsating white dwarfs that
are several years past outburst with past results on six other systems
shows that the instability
strip for accreting pulsating white dwarfs 
ranges from 10,500-15,000K, a wider range than evident for
ZZ Ceti pulsators. Analysis of the UV/optical pulsation properties
reveals some puzzling aspects. While half the systems show 
high pulsation amplitudes in the UV compared to their optical counterparts,
others show UV/optical amplitude ratios that are
less than one or no pulsations at either wavelength region. 

\end{abstract}


\section{Introduction}

In the decade since the white dwarf in the cataclysmic binary system 
GW Lib was found to show non-radial pulsations (Warner \& van Zyl 1998),
a dozen more systems of this type have been discovered (Warner \& Woudt 2004;
Wouldt \& Warner 2004; Araujo-Betancor et al. 2005a; Vanlandingham et al. 2005;
Patterson et al. 2005a,b; Mukadam et al. 2007; G\"ansicke et al. 2006;
Nilsson et al. 2006; Patterson et al. 2008; Pavlenko 2009). For
convenience, we will refer to the objects found from the Sloan Digital
Sky Survey as SDSShhmm$\pm$deg i.e. SDSS0745+45. The work on
all the known objects has allowed some progress
toward understanding how accretion affects the instability zones.
While normal non-interacting white dwarfs with hydrogen atmospheres
(DAVs or ZZ Ceti stars) show pulsations if they have temperatures in
the range of 10,800-12,300K with some dependence of the temperature range
on log g (Koester \& Holberg 2001; Bergeron et al. 2004; Gianninas et al. 2006; Mukadam et al. 2006; Castanheira et al. 2007),
the white dwarfs in cataclysmic variables (CVs) are known to be heated
by accretion (summaries in Sion 1991; 1999; Townsley \& G\"ansicke 2009).
Moreover, CV white dwarfs typically rotate an order of magnitude more rapidly
than single DA pulsators, and have solar composition or subsolar metal
composition
accreted atmospheres.
With very few exceptions, the temperatures of non-magnetic white dwarfs in
CVs
are above 12,000K and so they were previously not expected to be ZZ Ceti 
type pulsators. 
However,
HST ultraviolet observations of GW Lib clearly showed high amplitude
pulsations as well as a high temperature 
(Szkody et al. 2002a). In addition, the best fit to the data occurred with
a 2-temperature model, with 63\% of the white dwarf at a temperature of
13,300K and the rest at 17,100K. It was not known if the dual temperature was
related to the pulsation or to
the presence of a hotter boundary layer where the accretion disk 
meets the white dwarf. 

The existence of the pulsations at such a
high temperature led to speculation
that the pulsations in GW Lib could be due to a higher mass
white dwarf (Townsley, Arras \& Bildsten 2004) since a shift in g of about a factor of 
10 can shift the blue edge of the H/HeI instability strip 
to hotter temperatures by about 2000\,K. They also
stated that the effects of 
accretion i.e. 
heavier elements in the atmosphere of the white dwarf, or a faster spin than
present in ZZ Ceti pulsators could affect their models. Arras et al. (2006)
accomplished a more detailed study that showed that the atmospheric composition
of the accreting white dwarf is significant for determining the location of the
instability strip. They discovered that accreting model white dwarfs with a
high He abundance ($>$\,0.38) would form an additional hotter instability
strip at $\sim$15,000\,K due to HeII ionization. 

However, the discovery of pulsations in V455 And (Araujo-Betancor et al. 2005a)
and a subsequent snapshot HST spectrum showed that 
this white dwarf was in the normal
ZZ Ceti instability zone (within the uncertainties),
 with T$_{wd}\sim$10,500K and log g=8, thus
very different from GW Lib. 
Arras et al. (2006) suggested that the differences in the two
objects could be due to differences in the mass of the white dwarf and to
the He abundance in the driving
zone. Temperatures determined for 
 three more accreting pulsators provided further confusion (Szkody
et al. 2007), as all three systems (SDSS0131-09, SDSS1610-01, SDSS2205+11)
showed white dwarfs with temperatures
of 14,500-15,000K. Thus, it was not clear if GW Lib or V455 And was
the normal case for pulsating white dwarfs in CVs. Further complicating
the issue was the fact that the other known cool non-magnetic white dwarf 
in a CV
 (EG Cnc) shows no evidence of pulsations even though its temperature is
12,300K (Szkody et al. 2002b) and there are several CVs with T$_{eff}$ near
15,000K as well that don't pulsate e.g. WZ Sge, BC UMa, SW UMa (Sion et al.
1995; G\"ansicke et al. 2005). The situation for CVs containing magnetic
white dwarfs is similar. Araujo-Betancor et al. (2005b) determined
temperatures for seven systems with magnetic white dwarfs and found a range of
10,800-14,200K but none are known to show any signs of pulsations.

Our past HST observations of GW Lib, SDSS0131-09, SDSS1610-01 and SDSS2205+11 
showed similar pulsation frequencies in the UV as
the optical, with increased amplitudes (6-17) in the UV over the optical.
This amplitude ratio is consistent with low order modes (Robinson et al.
1995). Each non-radial pulsation mode can be described with a unique set
of indices. Mode identification is essential in asteroseismology to determine
the inner structure of the stars. The non-radial luminosity variations observed
in white dwarfs are almost entirely due to temperature fluctuations (Robinson,
Kepler \& Nather 1982). The stellar surface is divided into zones of higher
and lower T$_{eff}$ depending on the degree of spherical harmonic $\ell$. 
Since the stellar disk can't be resolved, geometric cancellation results
in smaller observable amplitudes for these modes. At UV wavelengths, the 
increased limb darkening decreases the contribution of zones near the limb,
so modes with $\ell$=3 are canceled less effectively in the UV compared to
the lower $\ell\leq$2 modes. But $\ell$=4 modes do not show a significant
change in amplitude as a function of wavelength. Robinson et al. (1995)
suggested a new mode identification technique based on this effect.
While the amplitude ratios for SDSS1610-01
 were consistent with those expected for an $\ell$=1
mode and T$_{eff}$=12,500K white dwarf, the derived temperature from
the spectral fit was 14,500K (Szkody
et al. 2007). In addition, long term optical coverage of GW Lib (van Zyl et al.
2004) and SDSSJ0131-09 (Szkody et al. 2007) showed that there is a high
degree of variablility in the amplitudes of the pulses so that at times,
some of the periods are not visible. This is normal behavior (termed
amplitude modulation) that is
observed in the cool ZZ Ceti stars 
 e.g. Kleinman et al. (1998), Mukadam et al. (2007).

In order to gain further insight into the location of the instability
strip for accreting white dwarf pulsators, and to constrain the mode
identification of the observed pulsation periods,
we obtained HST and nearly simultaneous optical observations of six other CV systems known
to contain pulsating white dwarfs (PQ And, REJ1255+26, SDSS1514+45,
SDSS1339+48, SDSS0745+45, SDSS0919+08 and SDSS1514+45). The basic
properties known for these six objects, along with the full SDSS names,
 are given in Table 1. Some
preliminary results appear in Mukadam et al. (2009).

\section{Observations}
\subsection{HST Ultraviolet Data}

The Hubble Space Telescope (HST) Solar Blind Channel (SBC) on the
Advanced Camera for Surveys (ACS) was used to observe each of the six
objects for 5 satellite orbits with either grating PR110L
or PR130L. While both gratings provide spectra from $\sim$1200\AA\ to
$\sim$2000\AA, the PR110L extends slightly bluer while the PR130L
has increased sensitivity near 1300\AA. In both cases, the prism
produces non-linear resolution,  with about 2\AA\ pixel$^{-1}$ (PR110L) and 
1\AA\ 
pixel$^{-1}$ (PR130L) at 1200\AA\ and about 40\AA\ pixel$^{-1}$ 
at 2000\AA. Since there is no time-tag mode for ACS, 60 or 61s
integrations times were used throughout 5 HST orbits on each source.
 With the setup time
during the first orbit, there were 134 or 138 integrations on each
object. 
The dead-time between integrations was 40s so the time resolution
is 100 or 101s. The initial exposure and centering
of the target was done with the F140LP filter with integration times
of 25-60s depending on the brightness of the source. 
 The observation
times and gratings are summarized in Table 2.

The HST data were analyzed with the reduction package aXe1.6 provided
by STScI. The targets were extracted with different widths that were
determined to optimize
the spectral and light curve results. For the spectra, a wide extraction
($\pm$17 pixels corresponding to $\pm$0.5 arcsec) was used to maximize the flux level, whereas a narrower
extraction of 5 (PQ And, REJ1255+26), 7 (SDSS0745+45, SDSS1514+45), 11 
(SDSS1339+48)
and 13 (SDSS0919+08) pixels was used for the light curves to optimize the 
best S/N for each system. 
To obtain light curves that could be analyzed for periodicity, all
the narrow extractions were summed over the useful wavelength range
 to obtain one UV flux point
per integration time interval.

The entire set of wide-extraction spectra during the five HST orbits were added
together to produce an average final spectrum for each object. These
average spectra are shown in Figure 1 ordered by increasing far-UV flux
from top to bottom. 
While the resolution is poor, the emission line of CIV (1550\AA) is apparent in
all systems. It is clear from this figure that there is a large
range in temperature as well as emission line flux for the six objects.

\subsection{Optical Data}

Due to the remote but possible chance of an outburst during the HST observations
 that could produce more
UV light than the limits of the ACS detector, each system was monitored
prior to and during each HST observation by amateurs (from the
American Association of Variable Star Observers) and professional
astronomers worldwide. These observations showed all six systems to
be close to quiescent values. The outburst history of the six objects
is given in Table 1. While PQ And and REJ1255+26 had published outbursts
in the literature, the previous outburst of SDSS0745+45 was only found
from the Catalina Real-time Transient Survey (Drake et al. 2009) which 
recorded an outburst with a minimum amplitude of 5 mag in 2006 October, with
the system brightness declining to its quiescent level over many 
months\footnote{click on object link in table at http://nesssi.cacr.caltech.edu/catalina/20050301/SDSSCV.html}.

Time-resolved ground-based observations as close in time as possible to the
HST UV observations were also coordinated in order to determine the amplitude
and period of optical pulsations that would aid in mode identification.
Seven observatories (Table 2) participated in providing observations.
The Apache Point Observatory (APO) 3.5m telescope was used with the
time-series photometric system Agile which uses a frame-transfer CCD 
and a BG40 filter to provide broad-band
blue light. 
The McDonald Observatory (MO) 2.1m telescope
with their time-series system Argos (Nather \& Mukadam 2004) 
and a BG40 filter was used in
a similar fashion. The 1m New Mexico State University (NMSU) telescope
with a CCD and BG40 filter and the 1m US Naval Observatory Flagstaff Station
(NOFS) 
telescope with a BG38 filter
were also used for several nights.
The 0.35m Schmidt-Cassegrain telescope equipped with an 
unfiltered SBIG CCD also provided data from
 the Sonoita Research Telescope (SRO) in
Arizona. In addition to the observations obtained close in time to the HST times,
further data on SDSS0919+08 (APO) and on PQ And (using the WIYN 3.5m telescope
equipped with OPTIC and a BG39 filter) were taken a year after the HST observations.
All photometric points  were made using differential
photometry with respect to comparison stars on the frames and light
curves were constructed using standard 
 IRAF\footnote{{IRAF (Image
 Reduction and Analysis
Facility) is distributed by the National Optical Astronomy Observatories, which
are operated by AURA,
Inc., under cooperative agreement with the National Science Foundation.}}
programs for sky-subtracted 
aperture photometry. To search for periodicities, a discrete
Fourier transform up to the Nyquist frequency was computed for each object, 
after first converting the light curves to
a fractional amplitude scale by dividing by the mean and then subtracting one.
A summary of the optical observations is also given in Table 2.

\section{Spectral Results}

Figure 1 shows that there is a large
range in continuum flux and shape 
as well as in the emission line flux of CIV (1550\AA)
 for the six objects. Since the core of Ly$\alpha$ does not reach zero
for any of the objects, there is an additional source of continuum light other
than the white dwarf, although this contribution is small in most cases.
This  ``second
component'' has been observed in UV observations of most quiescent
dwarf novae, and has been modelled in terms of a hot boundary layer
(e.g. Long et al. 1993) or an accretion belt
(e.g. G\"ansicke \& Beuermann 1996; Sion et al. 1996).
Long et al. (2009) showed that the WD
parameters inferred from a composite fit depend only very mildly on
the details of the model assumed for the second component.

 The CIV (and to a lesser
extent CII at 1335\AA, SIV at 1400\AA\ and NV at 1240\AA) emission lines
probably originate in the disk chromosphere. There may also be some
contribution from L$\alpha$ from the disk, although the large upturn at
Ly$\alpha$ in Figure 1 for SDSS1514, SDSS1339 and SDSS0919 is likely
geocoronal. 
From our past work on CV systems (e.g. G\"ansicke et al. 2005), we have
derived a procedure to pull out the temperature of the white dwarf.
 We account for
the disk continuum contribution by adding in a component whose flux
raises the continuum in the core of Ly$\alpha$ to the value observed.
As our objects were observed during (or close to) 
dwarf nova quiescence, our current
understanding of the theory of dwarf nova outbursts suggests that the accretion
disk should be very cool and optically thin. 
Since we do not know the spectral shape of an optically thin accretion disk, we try
three possibilities: a black body, a power law and a constant flux level
in F$_\lambda$. While these possibilities do not model line features from a
disk, the disk contribution is generally small in all cases.
Finally, we use simple broad Gaussians at the positions of the known UV lines
to fill in their contribution. For each disk contribution type, we match
the spectrum with a grid of Hubeny white dwarf models (Hubeny \& Lanz 1995).
The ACS data do not allow us to infer the surface gravity and
we hence fix it at $\log g=8.0$, as well as the metal abundances at
0.01 times their solar value. Assuming $\log g$ higher (lower) by
0.5\,dex would result in best-fit temperatures lower (higher) by
$\sim1000$\,K, which should be considered as a systematic error of our
method.

The results of our fits are listed in Table 3. The final
resulting temperature of the white dwarf is indeed almost independent
on the assumed spectral shape of the second component, but the 
exercise provides
an estimate of the error of our fits (1000K). If one expects a wildly varying 
white dwarf mass
distribution among these objects (which is not suggested by the
observations, e.g. Littlefair et al. 2008), the two
uncertainties related to the unknown white dwarf mass and the nature
of the second component should be added in quadrature. 

For completeness, we also 
include in the Table the
results for the 3 objects observed with the SBC in the past 
(Szkody et al. 2007)
that were re-analyzed with the new aXe1.6 software and refit in the same
manner as for our six new observations.
An example of the
fits for the three spectral shapes of the disk are shown for SDSS1339+48
in Figure 2 while the fitting results using a black-body for the disk 
contribution are
shown for all six of our new objects in Figure 3.

As a check on the white dwarf temperature, the $g$ magnitude for the
resulting white dwarf model is also listed and compared to the observed
SDSS photometric values in Table 3. All the white dwarf values are
fainter than observed, which is reasonable given that the accretion disk
will have some contribution to the optical light above that of the white
dwarf.

For a further exercise in the total spectral fit, we combined the
HST data with the available SDSS spectra for 5 of our objects. REJ1255+26
only has SDSS photometry as a spectrum was not obtained. For this object,
the $ugriz$ magnitudes were converted to fluxes. The  disk
component (using the constant distribution) was then subtracted from
the SBC spectrum and the result plotted with the optical spectrum. Figure 4
shows the combined UV and optical data for each system along with
the best fit white dwarf temperature models within $\sim$2000K. These
plots show the goodness of our temperature fits as well as the amount
of the disk contribution to the optical flux (the excess seen in the
observed fluxes over the model white dwarf).

\section{Light Curves}

We computed DFTs of all the HST and available optical light curves, and used
linear and non-linear least squares analyses to determine periods, amplitudes,
and phases of any coherent variability present in the data.
White noise was determined
by a light-curve shuffling technique where each time value of fractional
intensity was randomly reassigned to another existing time value. This
shuffling destroys coherent frequencies but keeps the same time sampling
and white noise as the original data. The DFT of the shuffled light curve
provides the noise at each frequency; the mean of the average noise from
10 shufflings is taken as the white noise of the light curve and 3 times this
value is quoted in our paper as a detection limit for pulsations. The
amplitudes are given in millimodulation amplitudes (mma) where 1 mma is
a 0.1\% change in intensity.
The DFTs for the HST data are shown in Figure 5 along with the window
functions. The window function is the DFT of a single frequency
noiseless sinusoid sampled exactly the same times as the actual light
curve. These functions look quite similar to each other as the HST
data for all six systems is sampled in almost the same way (since we
deleted data points when the background noise was too high, the sampling
is not completely identical for all six targets). A summary of the
observed periods and limits from the HST
and optical data is contained 
in Table 4, and the details for each system are discussed
below.

\subsection{PQ And}

For this system, the ground and space observation time overlaps for 4 hrs (Table 2),
so periods and amplitudes can be optimally compared. 
Figures 5 and 6 show the DFTs for 
the UV and optical data obtained on 2007 Sept. 13. The ground
data were obtained with a smaller telescope but the observation interval
is longer so that the mean noise is lower (5 mma) than for the UV
data (14 mma). While the optical
data show the presence of two periods (2337 and 1285s) 
with amplitudes near 20 mma,
there is no signifant pulsation evident in the UV. This lack of UV
pulsations is surprising and without explanation. The 1285s optical
period is the
pulsation period reported by Vanlandingham et al. (2005) 
(Table 1)
and the data on 2007 Sept. 15 and 17 show a similar
period within the errors. Thus, we regard this as the non-radial pulsation. The 2337s
period is not repeated on the other dates and 
is likely due to flickering noise. 

Since the four
previous systems observed with HST had shown pulsation amplitudes at
least 6 times higher in the UV than the optical, consistent with $\ell$=1 or 2
modes, we would expect a UV amplitude of at least 60 mma at the 1285s
period that was evident in the optical within the same time interval. Thus, 
the lack of detection of any pulsation in the UV implies a high
$\ell$ mode in this system, if the optical periodicity at 1285s is a non-radial
pulsation.  High $\ell$ modes ($\ell>$2) have never been clearly
and unambiguously identified in the ZZ Ceti stars. For example, Thompson et al. 
(2004)
identified the 141.9\,s period in PY Vul (G185-32) as an $\ell$=4 mode, but Pech et al.
(2006) suggest that it is an $\ell$=2 mode. Furthermore, they explain that the
low value of UV/optical amplitude could be attributed to a resonance between the
$\ell$=2 mode and nearby $\ell$=3 and $\ell$=5 modes, which remain undetectable.

  The high quality WIYN data obtained a year after the HST 
observation show an even higher amplitude new periodicity at 679s along 
with its sub-harmonic at 1355s (Figure 7). Harmonics and linear 
combinations in our data could arise as a result of non-linearities 
introduced by relatively thick convection zones (Brickhill 1992, Brassard 
et al. 1995, Wu 2001, Montgomery 2005). However, since we observe a 
sub-harmonic at 2P\,=\,1355s and do not detect a harmonic at 
P/2\,=\,339.5\,s, we can rule out convection as the likely cause of the 
observed non-linearity. Since there is only one observation of this new 
periodicity, it is possible that both the 679s period and its sub-harmonic 
at 1355s were caused by flickering. Should frequent monitoring of the 
system fail to reveal the 679s period again, then the 
flickering hypothesis is likely. If however the 679s mode proves to be persistent, 
then it might be an unusual pulsation mode with an 
observed sub-harmonic instead of a typical harmonic. As a pulsation mode, 
we could explain its appearance 
and the absence of previously observed modes as due to amplitude modulation. 
Amplitude modulation is apparent in GW Lib (Van Zyl et al. 2004) and 
SDSS0131-09 (Szkody et al. 2007) and also the ZZ Ceti G29-38 (Kleinman et 
al. 1998). We currently favor the idea that the 679s mode 
and its sub-harmonic at 1355s are caused by flickering.

\subsection{SDSS0745+45}

The HST observations of this system took place in the midst of
a ground-based run of several nights at McDonald Observatory and
SRO. Both observatories overlapped with HST times 
for about 2.5 hrs (Table 2). While the light
curve is highly modulated in both the UV and optical, the modulation is 
at the long period of 86 min (and its 2nd harmonic at 43 min and 3rd harmonic
at 28 min),
not within the  previously observed range of pulsation periods between
 1166-1290s that were clearly evident during
 7 nights from 2005 Oct through 2006 Jan
(Mukdadam et al. 2007). 
Figure 8 shows the intensity curves and DFTs for the 2007 data in
comparison to that of 2006. While the pulsation is clearly evident
in the light curves in 2006, it disappeared at the time of the HST
observation. Combining the 4 nights of optical data from 2007
Oct 30-Nov 2, we obtain a 3$\sigma$ limit to the pulsation amplitude
of 8.5 mma. The same limit from the HST data is 12 mma (Figure 5). 

Further complicating the issue of the disappearance of the pulsation is
the origin of the 86 min modulation. This period (or its harmonics) was
evident in 2 of the 7 nights in 2005-2006 where it was presumed to
be the orbital period (Mukadam et al. 2007). However, recent spectroscopy
(John Taylor 2009, private communication) has revealed an orbital period
of 77.8$\pm$1.5 min. Combining the 4 nights of photometry listed in
Table 4, we obtain periods of 89.3$\pm$0.02 min and 43.23$\pm$0.01 min.
Averaging these numbers (weighted inverseley as the squares of the
uncertainties), we obtain a photometric period of 87.4$\pm$1.3 min. Clearly, the
spectroscopic period is significantly less than the photometric period.
Photometric periods that are 2-4\% longer than spectroscopic periods
are usually ascribed to superhumps, a phenomenon usually observed
following outbursts in short period systems due to precession of an
eccentric disk caused by the heating from the outburst (Warner, 1995;
Patterson 2001). While it is unusual to have a superhump present at
quiescence, one is also apparent in V455 And (Araujo-Betancor et al. 2005a).
It is perhaps even more unusual that the periods are so different in
SDSS0745+45 (12\%). Even if
the error bars are underestimated by a factor of 2, 
the difference between the spectroscopic and photometric periods
is still 5\%. While this difference is 2.8\% in V455 And, there is a
one-day alias in the photometric period so the difference could possibly be as
large as 9.2\%, similar to what we find for SDSS0745+45.

As previously noted, 
the detection of an outburst around mid-Oct 2006 
from the Catalina Sky Survey data means it is likely 
that the white dwarf in SDSS0745+45 
had not completely cooled to its quiescent temperature by the time of the HST
observation. Indeed, this object is the hottest one of all the objects we have
observed, which is consistent with it having been heated during the outburst.
Past work on the cooling times of white dwarfs following
outbursts e.g. WZ Sge (Slevinsky et al. 1999; Long et al. 2003;
Sion et al. 2003; Godon et al. 2006), and AL Com (Szkody et al. 2003) have shown that this
cooling can take more than 3 years. The outburst could have caused some increase in the 
eccentricity of the disk 
that causes the long period modulation to be more prominent than pre-outburst. 
We expect the white dwarf to
continue to cool during the next two years and the pulsations to
resume, while the long period modulation decreases. 

\subsection{SDSS0919+08}

Whereas five out of six past optical light curves of this object from
2005 Dec to 2007 Mar showed
a period near 260s with amplitudes of 7-16 mma (Mukadam et al. 2007), Figure 5 
shows that the HST data in 2007 Nov only reveals
a harmonic of the orbital period at 40.75 min, with a limit of 15 mma to
any shorter term periodicity in the UV (the two peaks on both sides of
the 40  min period are aliases). Optical data obtained
2 days prior to the HST observations also reveal no period to a limit
of 4 mma (Figure 10) and the data at the end of 2008 show no pulsation to a limit
of 7 mmma (Table 4). Since
there are no known outbursts of this system, and the white dwarf
temperature is not high, it is not at all clear why the white dwarf in
this system has stopped pulsating. Mukadam et al. (2007) thought the 260s
period was a close doublet and the one night it was not observed in their
data could
be a beating of the two frequencies. However, the lack of periods on
the three separated days of observations in our data and especially the lack
of pulsations in the UV indicates this system has actually stopped pulsating.
It is possible that if this system is close to the edge of its instability strip,
accretion related heating could push it outside. However, the amount of accretion
would need to be small enough that it would not noticeably affect the visual
magnitude of the system.

\subsection{SDSS1339+48}

From three nights of photometry in 2005, G\"ansicke et al. (2006) identified
a prominent pulsation at a period of 641s, as well as a long period at
either 320 or 344 min but no modulation at the spectroscopically determined
period of 82.52 min. Neither our HST data (Figure 5) nor our ground-based
photometry obtained two days prior to the HST data (Figure 10) 
show the pulsation period
of 641s. The HST data do show one significant long period at 7.4 hrs
and two short periods (210.3 and 229.6s) that are close to the 
Nyquist
frequency for the data resolution but just below the 3$\sigma$ noise level
of 53 mma. The optical data reveal none of these
periods but do show a period consistent with the orbital period (within
the error bars), as well as a period at 1539s. Due to the limited data,
it is difficult to conclude that any of these periods result
from anything other than flickering or accretion effects. However, it
is clear that there is no large amplitude UV pulsation at the previously
determined optical period of 641s .

\subsection{SDSS1514+45}

 Our optical data from APO (Figure 11) have about 1.5 hrs of overlap with the HST UV
data for this source. Neither wavelength shows the 
559s period previously 
reported by Nilsson et al. (2006). As that period has not been
evident since their data in 2005, further observations are needed to
confirm if this really is a pulsator with a changing amplitude.

The DFT from the UV data (Figure 5) does show a longer period at 88.8 min 
which could be the orbital period. The faintness of this system has
precluded the determination of an orbital period from spectroscopy at
the current time. There is a period near 700s that is almost at the
3$\sigma$ limit, but as this is not evident in the optical, it is
likely related to flickering.

\subsection{REJ1255+26}

Our APO optical time-resolved data on this system only encompass one light
curve on the night following the HST observations. The UV data show the
1.9 hr orbital period but a limit of 52 mma to any shorter periods.
The optical data (Figure 13) reveals two short periods at 654.5 and 582.1.
Patterson et al. (2005b) had
previously found a period of 668s. As in
PQ And, the presence of optical periods without detection in the UV at
a higher amplitude implies a high $\ell$ mode of pulsation, if the
periods are due to non-radial pulsations.

\section{Discussion}

Figures 1,3,4 and Table 3 show that the 9 systems with similar SBC
data show a range of temperatures from 10,000-17,000K. The hottest
temperature white dwarf exists in SDSS0745+45. However, since that system
underwent an outburst only 1 year prior to the HST observation, it
is likely that its white dwarf has not cooled to its quiescent value.
This is corroborated by the lack of pulsations evident in the HST
data. Long term studies of the white dwarf cooling in
low accretion rate CVs with short orbital periods have shown that it takes
more than 3 yrs for the white dwarf to cool following outburst 
(Piro et al. 2005; Godon
et al. 2006). Eliminating SDSS0745+45 and adding in the
temperatures derived from STIS observations of GW Lib (15,400 for
a black body disk contribution; Szkody et al. 2002a) and V455 And
(11,500; Araujo-Betancor et al. 2005a), and from the eclipse modeling of 
SDSS1507+52 (11,000; Littlefair
et al. 2008), gives an instability range of 10,000-15,000K
with the CV pulsators spread throughout this range. Thus, if the higher
temperatures of pulsating accretors relative to ZZ Ceti stars is due to
mass or composition, then no unique parameter characterizes the accreting
pulsators. 
 Half of the 11 temperatures lie within the ZZ Cet
instability strip (within the error bars and with a log g of 8) and half are 
hotter. Figure 13 shows our 11 quiescent
temperatures along with the ZZ Ceti empirical instability strip (Gianninas
et al. 2007). While a few of the systems could fall within the strip if
they have massive white dwarfs (higher log g), it would be difficult to
have mass be the primary cause of the width of the strip. Arras et al. (2006)
can account for increased width with an increase in He in the driving zone.

Besides the width of the instability strip for accreting pulsators, the
other oddity is why all CVs with white dwarfs
in the temperature range of the known pulsators do not show pulsations.
Table 6 lists all the white dwarfs with reliable temperature determinations
(from the recent summary in Townsley \& G\"ansicke 2009, with the addition
of SDSS1507+52 from Littlefair et al. 2008) and they are also
plotted in Figure 13. The two closest systems to the ZZ Cet instability strip 
are EG Cnc (at the blue edge) and SDSS1035+05 (at the red edge). Further 
monitoring of these systems is
needed to determine if they never pulsate or if they were observed during a
hiatus of their pulsations. 

GW Lib still stands out as the only system among
the 11 with available UV spectra in which the best fit is obtained
with a dual temperature white dwarf rather than a white dwarf plus
an accretion disk. It is the only system in which the core of Ly$\alpha$
does reach zero. If this dual temperature is due to a boundary layer
ring that is hotter due to the accretion, it is not clear why this is
not the case in the other 10 systems, especially for SDSS0745+45 which
is heated by the recent outburst.

Within the limited numbers, there does not seem to be any correlation of
white dwarf temperature
with orbital period as would be expected if the white dwarfs are
cooling due to decreasing mass accretion rate
 as they evolve to shorter orbital periods
(Howell, Nelson \& Rappaport 2001).
 Figures 2-4 give some indication
of the accretion luminosity (hence white dwarf mass and accretion rate)
from the contribution of the accretion disk to the
light. Excluding SDSS0745+45, whose white dwarf is likely not at its
quiescent temperature, the white dwarfs in our models contribute 75-89\% of
the UV light and 42-75\% of the optical light for the other 8 objects in
Table 3.
The three hottest white dwarfs (SDSS2205+11, SDSS0131-09 and SDSS1610-01)
do have the
three highest disk contributions to the optical light (50, 58 and 41\%)
which is consistent with a higher mass white dwarf or a higher accretion
rate that would heat the white dwarf. 

Figure 14 shows the accreting pulsators
as well as the non-pulsating white dwarfs (Table 6) as a function of
their orbital period. As noted in G\"ansicke et al. (2009b), most of
the pulsating white dwarfs lie in a period regime termed the "period spike"
between 80-86 min. The most egregious outlier is REJ1255+266 near 2 hrs. 
Since the period determination for REJ1255+266 is a photometric one
and we now know that at least 2 of our pulsating accretors (V455 And and
SDSS0745+45) show different
photometric periods than spectroscopic, this period may be suspect and
awaits a spectroscopic determination.
However, the pulsator SDSS1507+52 has a period of 66.6 min determined from eclipses
(Patterson et al. 2008) so it appears this one really is an
outlier to the rest of the group. GW Lib and SDSS0745+45 (not shown on
the plot as its temperature may not be its quiescent one) are outside the
spike as well. It is clear from Figure 14 that most of the accreting pulsators
are concentrated in a narrow period range near 82 min. It is easy to
explain why longer period
systems are not evident as pulsators, as those
typically have increased mass accretion rates which can hide the
white dwarf, and thus make pulsations harder to detect. But it is not obvious
why the pulsations would not be present in systems with shorter orbital  periods and why the 
transition is
so abrupt.

A pure instability strip with non-pulsators outside and only pulsators 
within, implies that pulsations are an evolutionary phase along the 
cooling track (Fontaine et al. 1982, 2003, 
Bergeron et al. 2004, Castanheira et al 2007). In other words, {\it all} accreting white 
dwarfs would undergo the pulsation phase. An impure instability strip with 
pulsators and non-pulsators mixed in implies that parameters other than 
the white dwarf temperature and mass are at play in deciding whether the 
star will pulsate or not (Kepler \& Nelan 1993, Mukadam et al. 2004). Our 
preliminary results indicate an impure strip with 13 non-pulsators in the 
range of 10500--15400\,K; this lends credence to the theory that He 
abundance is the third parameter that decides the shape of the instability 
strip for accretors (Arras et al. 2006).

In addition to the temperature data, we now also have ten systems with
pulsation properties determined from UV through optical wavlengths
(Table 5). This table shows a perplexing mix of results. While
several objects (GW Lib, SDSS0131-09, SDSS1610-01 and SDSS2205+11)
behave similarly to ZZ Ceti stars in having high UV/optical amplitude
ratios, and similar periods evident in UV and optical, others
(PQ And and REJ1255+26) have very low UV/optical amplitude ratios
or show no pulsations in either UV nor optical at the time of our
observations (SDSS0745+45, SDSS0919+08, SDSS1339+48, SDSS1514+45).
The absence of pulsations in SDSS0745+45 can be explained
by its recent outburst, which has likely heated its white dwarf and moved
it out of the instability strip. This explanation is corroborated by
the lack of optical and UV pulsations of GW Lib following its recent 
outburst (Szkody et al. 2009; Copperwheat et al. 2009). SDSS1514+45 
suffers from insufficient data to
confirm pulsations.  However, there is no clear explanation for SDSS0919+08 and
SDSS1339+48. Recent observations of SDSS2205+11 (Southworth et al. 2008)
also show a lack of pulsation on two nights as compared to previous 
years (Warner \& Woudt 2004; Szkody et al. 2007).
While SDSS1339+48 is just outside the blue edge of the ZZ Cet instability
strip for log g=8 (Figure 13), which might account for its lack of pulsation
if its white dwarf underwent a slight temperature increase, SDSS0919+08
and SDSS2205+11 are much further away from this edge.
 Southworth et al. (2008)
discuss several possible reasons for the lack of pulsations (destructive
interference, changes in the thermal state of the driving region, or
changing visibility of the modes over the surface of the white dwarf)
but conclude further observations over long timescales are needed to sort
out the cause. While different frequencies are known to be dominant
at different times during the course of several years in GW Lib (van Zyl et al.
2004) and in SDSS0131-09 (Szkody et al. 2007), information as to the
length of time that systems show no pulsations at all is not available.

\section{Conclusions}

Our recent ultraviolet and optical data on six systems combined with
past data on six others 
produces a dataset of 12 accreting pulsating
white dwarfs with
temperatures (Tables 3, 6) and ten with 
pulsation properties determined from UV through
optical wavelengths (Table 5). These datasets are beginning to contain
enough objects to begin to see some trends as well as to reveal some
abnormalities.

\begin{itemize}
\item The temperature range for the 11 objects that have been in quiescence
for years are in the range of 10,500-15,000K. This range includes the
temperatures found in ZZ Ceti pulsators but extends to higher
temperatures. The temperatures
appear to uniformly spread throughout the entire range, but we are still in the domain of
small number statistics. Since we do not have masses or He abundances for
these 11 systems, it is difficult to test the predictions of
(Arras et al. 2006) as to the cause of the width of the strip.

\item Several CVs are known to have white dwarfs with temperature within this
range of 
10,500-15,000K but they are not pulsating. Table 6 presents a list of
the white dwarfs in disk-accreting CV that have reliable quiescent 
temperatures (from Townsley \& G\"ansicke
2009, Littlefair et al. 2008, and this work) and whether they have shown 
pulsations.
It is unclear why not all CVs in this temperature range are pulsating.
In addition to the disk-accretors, there are an additional 11 highly magnetic
white dwarfs in CVs (Polars) in the Townsley \& G\"ansicke (2009) list
within this temperature range, none of which show pulsations.

\item GW Lib is the only system among the 11 with UV spectra in which
the best fit is obtained with a dual temperature white dwarf rather than
a white dwarf plus a disk contribution.

\item The object that is one year past outburst (SDSS0745+45) has a
hotter temperature (17,000K) and is no longer pulsating. This corroborates
past work that shows the white dwarfs in CVs are heated for several
years following outburst (Piro et al. 2005; Godon et al. 2006) and it is likely that this
object has moved out of the instability strip. 

\item  Four systems
(GW Lib, SDSS0131-09, SDSS1610-01 and SDSS2205+11) show identical
pulsation periods in the UV and optical with high amplitude ratios of
UV/opt. These four are similar to the low order modes evident in
ZZ Cet stars; however, the temperatures of all 4 are 14,000-15,000,
far outside the range for ZZ Ceti objects.  

\item Two objects (PQ And and REJ1255+26) have UV/optical amplitude 
ratios that
are less than 1. Both of these objects have white dwarf temperatures
of 12,000K, just inside the blue edge of the ZZ Ceti instability strip (Figure
13). If the
optical periods observed are non-radial pulsations, this implies a
high $\ell$ mode not unambiguosly observed in ZZ Ceti pulsators. It remains
a problem for theorists to determine if the differences between accreting
and non-accreting white dwarfs could cause this effect. 

\item Four objects (SDSS0745+45, SDSS0919+08, SDSS1339+48 and SDSS1514+45)
did not show pulsations in either UV nor optical. The lack of pulsations
in SDSS0745+45 can be explained by its outburst only 1 year prior to the
HST data, while SDSS1514+45 has had only minimal observational data in the past
(Nilsson et al. 2006), so further data are needed to confirm it as an accreting pulsator.  For the other 2 systems, there is no clear explanation for why
the pulsations that were evident from several nights of ground-based data
in the past have disappeared. Both of these objects are in the intermediate
temperature zone (12,500-13,500K) between the normal ZZ Ceti instability
edge and the hotter temperatures of the 4 pulsators that do show UV
pulsations of high amplitude. Since the temperatures are not high (as in
SDSS0745+45), it is unlikely that an outburst has occurred in these 2
systems. It is possible that these pulsators were close to the edge of
the instability strip (SDSS0919+08 would have to be high mass to be near
the edge), and changes in accretion heating take them in and
out of the instability zone. We intend to keep observing these objects to
determine if pulsations return at the previous periods. 

\item Two objects (V455 And and SDSS0745+45) show photometric periods
that are longer than the spectroscopic periods by 3-12\%. The presence
of such periods are typically due to superhumps caused by precessing,
eccentric disks in short orbital period CVs during an outburst. If
this is the cause in these two objects, it is unusual to find 
the superhumps during quiescence and even more unusual to have such
a large percentage difference in the periods.
\end{itemize}

To make progress toward understanding the edges of the instability strip
and the pulsation modes that relate to the appearance/disappearance
of periods at different times, further data are needed. Mass determinations
of a few of the hot vs cool white dwarfs can determine how this
parameter effects the width of the strip. Continued observations of
the systems that have undergone outbursts in 2006-2007 (SDSS0745+45,
GW  Lib, V455 And and SDSS0804+51) as they re-enter the instability
strip following the heating during the outburst will provide valuable
information about the modes and depth of heating. Long observation
sequences for the objects in Table 6 that are not known to pulsate are
needed to place stringent limits on the amplitudes of possible pulsations.
 The identification
of further accreting pulsating systems will help to enlarge the database
on which to draw conclusions about the stability of periods and the behavior
as a function of the temperature of the white dwarf.

\acknowledgments

We gratefully acknowledge the many amateur and professional 
observers who monitored our objects
prior to and during the HST observations which allowed the UV observations
to proceed with the knowledge that the objects were at quiescence. Special
thanks go to Agatha Raup, Joanne Hughes and Gary Walker.
This research was supported by NASA grant HST-GO-11163.01-A from the Space
Telescope Science Institute which is operated
 by the Association of Universities for Research in Astronomy, Inc., for NASA,
 under contract NAS 5-26555.
BTG was supported by a PPARC Advanced Fellowship.

\clearpage
\begin{deluxetable}{lccllll}
\tablewidth{0pt}
\tablecaption{Objects Observed with HST}
\tablehead{
\colhead{Name} & \colhead{Mag} & \colhead{P$_{orb}$(min)} &
\colhead{Opt Pulse P (sec)} & \colhead{Amp (mma)} & \colhead{Outbursts (yr)} & \colhead{Ref\tablenotemark{a}} }
\startdata
PQ And & 19.1(V) & 80.6 & 1263, 1286 & 25  & 1938, 1967, 1988 & 1,2,3 \\
SDSS0745\tablenotemark{b} & 19.0(g) & 77.8 & 1166-1290 & 45-70 & 2006 & 4 \\
SDSS0919 & 18.2(g) & 81.3 & 260 & 7-16 & none & 4 \\
SDSS1339 & 17.7(g) & 82.5 & 641 & 25 & none & 5 \\
SDSS1514 & 19.7(g) & ... & 559 & 12 & none &  6 \\
REJ1255\tablenotemark{b} & 19.1(g) & 119.5 & 668, 1236, 1344 & 7-30 & 1994 & 7 \\
\enddata
\tablenotetext{a}{1 Schwarz et al. (2004), 2 Patterson et al. (2005a),
3 Vanlandingham et al. (2005), 
4 Mukadam et al. (2007), 5 G\"ansicke et al. (2006), 6 Nilsson et al. (2006),
7 Patterson et al. (2005b)}
\tablenotetext{b}{Full names of objects are: SDSS J074531.92+453829.6;
SDSS J091945.11+085710.0; SDSS J133941.11+484727.5; SDSS J151413.72+454911.9;
RE J1255+266} 
\end{deluxetable}

\begin{deluxetable}{ccllrl}
\tabletypesize{\scriptsize}
\tablewidth{0pt}
\tablecaption{Summary of Observations}
\tablehead{
\colhead{Name} & \colhead{Obs} & \colhead{Instr} & \colhead{Filter} & \colhead
{Int(s)} & \colhead{UT Time} }
\startdata
PQ And & HST & SBC &  PR110L & 60x138 & 2007 Sep 13 03:36-10:45 \\
PQ And & NOFS & CCD & none & 180 & 2007 Sep 13 06:43-12:18 \\
PQ And & NMSU & CCD & BG38 & 180 & 2007 Sep 13 05:05-05:51 \\
PQ And & NMSU & CCD & BG38 & 180 & 2007 Sep 15 04:33-09:41 \\
PQ And & NMSU & CCD & BG38 & 180 & 2007 Sep 19 06:59-11:04 \\
PQ And & WIYN & OPTIC & BG39 & 25 & 2008 Oct 21 08:14-10:57 \\
SDSS0745 & HST & SBC & PR110L & 61x138 & 2007 Nov 01 02:48-09:57 \\
SDSS0745 & MO & Argos & BG40 & 30 & 2007 Oct 30 07:13-12:31 \\
SDSS0745 & MO & Argos & BG40 & 30 & 2007 Oct 31 07:05-12:33 \\
SDSS0745 & MO & Argos & BG40 & 30 & 2007 Nov 01 07:21-12:26 \\
SDSS0745 & MO & Argos & BG40 & 30 & 2007 Nov 02 07:17-12:34\\
SDSS0745 & SRO & CCD & none & 1200 & 2007 Oct 31 07:32-12:25 \\
SDSS0745 & SRO & CCD & none & 900 & 2007 Nov 01 07:27-12:27 \\
SDSS0919 & HST & SBC & PR130L & 61x134 & 2007 Nov 14/15 18:36-01:41 \\
SDSS0919 & MO & Argos & BG40 & 15 & 2007 Nov 12 10:13-12:43  \\
SDSS0919 & NMSU & CCD & BG39 & 180 & 2007 Nov 8-15 2 per night \\
SDSS0919 & SRO & CCD & none & 300 &  2007 Nov 14 08:55-12:37 \\
SDSS0919 & SRO & CCD & none & 300 &  2007 Nov 15 09:05-10:18 \\  
SDSS0919 & APO & Agile & BG40 & 20 & 2008 Dec 30 10:10-13:28 \\
SDSS1339 & HST & SBC & PR130L & 61x138 & 2008 Jan 25 07:42-14:51 \\
SDSS1339 & APO & Agile & BG40 & 15 & 2008 Jan 23 11:26-12:57 \\
SDSS1514 & HST & SBC & PR130L & 60x138 & 2008 May 08 05:24-12:29 \\
SDSS1514 & APO & Agile & BG40 & 40 & 2008 May 08 03:00-07:00 \\
SDSS1514 & APO & Agile & BG40 & 40,80 & 2008 May 09 04:53-07:04 \\
REJ1255 & HST & SBC & PR110L & 61x134 & 2008 May 14 01:50-08:58 \\
REJ1255 & APO & Agile & BG40 & 40 & 2008 May 15 03:00-07:02 \\
\enddata
\end{deluxetable}

\begin{deluxetable}{lcclc}
\tabletypesize{\scriptsize}
\tablewidth{0pt}
\tablecaption{Model Fits to SBC Spectra}
\tablehead{
\colhead{Obj+Model} & \colhead{T$_{wd}$(K)} & \colhead{d(pc)} & 
\colhead{T$_{BB}$/PL\tablenotemark{a} } & \colhead{$g_{wd}$} }
\startdata
SDSS1514+const & 10,500 & 408 & ... & 20.0 \\
SDSS1514+BB & 10,000 & 358 & 9500 & 19.9 \\
SDSS1514+PL & 10,500 & 416 & 1.0 & 20.1 \\
PQ And+const & 12,000 & 340 & ... & 19.3 \\
PQ And+BB & 12,000 & 361 & 16,500 & 19.4 \\
PQ And+PL & 12,000 & 337 & -0.16 & 19.2 \\
REJ1255+const & 12,000 & 403 & ... & 19.6 \\
REJ1255+BB & 12,000 & 543 & 11,500 & 20.3 \\
REJ1255+PL & 12,000 & 420 & 0.90 & 19.7 \\
SDSS1339+const & 12,500 & 191 & ... & 17.9 \\
SDSS1339+BB & 12,500 & 187 & 25,000 & 17.9 \\
SDSS1339+PL & 12,500 & 187 & 0.08 & 17.9 \\
SDSS0919+const & 13,500 & 319 & ... & 18.9 \\
SDSS0919+BB & 13,500 & 333 & 15,500 & 19.0 \\
SDSS0919+PL & 13,500 & 330 & 1.0 & 19.0 \\
SDSS0131+const & 14,000 & 388 & ... & 19.3 \\
SDSS0131+BB & 14,500 & 432 & 14,000 & 19.4 \\
SDSS0131+PL & 14,000 & 371 & 0.66 & 19.2 \\
SDSS1610+const & 14,000 & 504 & ... & 19.8 \\
SDSS1610+BB & 14,500 & 611 & 13,500 & 20.2 \\
SDSS1610+PL & 14,500 & 562 & 1.0 & 20.0 \\
SDSS2205+const & 14,000 & 859 & ... & 21.0 \\
SDSS2205+BB & 15,000 & 919 & 11,000 & 21.0 \\
SDSS2205+PL & 14,000 & 862 & 0.66 & 21.0 \\
SDSS0745+const\tablenotemark{b} & 17,000 & 445 & ... & 19.2 \\
SDSS0745+BB & 17,000 & 474 & 19,000 & 19.3 \\
SDSS0745+PL & 17,000 & 463 & -0.02 & 19.3 \\
\enddata
\tablenotetext{a}{T$_{BB}$ is the black body temperature of the 2nd component 
and PL is the slope of the power law for the 2nd component.}
\tablenotetext{b}{As this object underwent an outburst one year prior to
the HST observation, the white dwarf is likely not yet at its quiescent temperature.}
\end{deluxetable}

\begin{deluxetable}{lllll}
\tabletypesize{\scriptsize}
\tablewidth{0pt}
\tablecaption{Summary of Observed Periods and 3$\sigma$ Limits}
\tablehead{
\colhead{Object} & \colhead{Wavelength (\AA)} & \colhead{Period} & 
\colhead{Amp (mma)} & \colhead{UT Date}}
\startdata
PQ And & 1200-1820 & ... & $<$42 & 09-13-07 \\
PQ And & optical & 1285$\pm$10s, 2337$\pm$10s & 22, 22 & 09-13-07 \\
PQ And & optical & 1309$\pm$13s & 28 &  09-15-07 \\
PQ And & optical & 1301$\pm$13s & 27 & 09-19-07 \\
PQ And & optical & 679.0$\pm$1.4s, 1355$\pm$16s  & 28, 10  & 10-21-08 \\
SDSS0745 & 1240-1910 & ... &  $<$12 & 11-01-07 \\
SDSS0745 & optical & 84.4$\pm$0.4min\tablenotemark{a}, 44.1$\pm$0.3min\tablenotemark{b} & 68, 30 & 10-30-07 \\
SDSS0745 & optical & 86.4$\pm$0.4min\tablenotemark{a}, 41.5$\pm$0.2min\tablenotemark{b}, 27.8$\pm$0.2 & 68, 36, 17 & 10-31-07 \\
SDSS0745 & optical & 91.1$\pm$0.8min\tablenotemark{a} & 70 & 11-01-07 \\
SDSS0745 & optical & 83.6$\pm$0.3min\tablenotemark{a}, 42.2$\pm$0.3min\tablenotemark{b}, 28.5$\pm$0.2min & 62, 22, 17  & 11-02-07 \\
SDSS0919 & 1226-1955 & 40.72$\pm$0.09min\tablenotemark{b}  & 85 & 11-14/15-07 \\
SDSS0919 & optical & ... & $<$4 & 11-12-07 \\
SDSS0919 & optical & ... & $<$7 & 12-30-08 \\
SDSS1339 & 1245-1955 & 7.39$\pm$0.25hr & 104  & 01-25-08 \\
SDSS1339 & optical & 83.2$\pm$3.0min\tablenotemark{a}, 1539$\pm$32s & 24, 14  & 01-23-08 \\
SDSS1514 & 1324-1935 & 88.8$\pm$min\tablenotemark{c}  & 108 & 05-08-08 \\
SDSS1514 & optical & ... & $<$6  & 05-07-08 \\     
REJ1255 & 1335-1885 & 1.91$\pm$0.04hr\tablenotemark{a} & 76 & 05-14-08 \\
REJ1255 & optical & 54.5$\pm$1.0min, 582.1$\pm$1.7s, 654.5$\pm$1.9s & 12, 13 & 05-15-08 \\
\enddata
\tablenotetext{a}{Orbital period or superhump period.}
\tablenotetext{b}{First harmonic of orbital or superhump period.}
\tablenotetext{c}{Likely orbital period.}
\end{deluxetable}

\clearpage
\begin{deluxetable}{lllc}
\tablewidth{0pt}
\tablecaption{Summary of UV and Optical Periods and Amplitude Ratios}
\tablehead{
\colhead{Object} & \colhead{UV Periods (s)} & \colhead{Opt Periods (s)} &
\colhead{UV/opt amplitude}}
\startdata
GW Lib & 648, 376, 236 & 646, 376, 237 & 6-17 \\
SDSS0131 & 213 & 211 & 6 \\
SDSS1610 & 608, 304, 221 & 608, 304, 221 & 6 \\
SDSS2205 & 576 & 575 & 6 \\
PQ And & ... & 1285 & $<$ 1 \\
REJ1255 & ... & 582, 655 & $<$ 1 \\
SDSS0745 & ... & ... & ... \\
SDSS0919 & ... & ... & ... \\
SDSS1339 & ... & ... & ... \\
SDSS1514 & ... & ... & ... \\
\enddata
\end{deluxetable}

\clearpage

\begin{deluxetable}{llcc}
\tablewidth{0pt}
\tablecaption{All WDs in Disk CVs with Temperatures of 10,000-15,000K}
\tablehead{
\colhead{Name} & \colhead{P$_{orb}$ (hr)} & \colhead{T$_{eff}$K} & 
\colhead{Pulsating}} 
\startdata
SDSS1507+52 & 1.11 & 11,000$\pm$500 & Y \\
GW Lib & 1.28 & 13,300+17,100 & Y \\
BW Scl & 1.30 & 14,800$\pm$900 & N \\
LL And & 1.32 & 14,300$\pm$1000 & N \\
PQ And & 1.34 & 12,000$\pm$1000 & Y \\
SDSS1610-01 & 1.34 & 14,500$\pm$1500 & Y \\
V455 And & 1.35 & 10,500$\pm$750 & Y \\
AL Com & 1.36 & 16,300$\pm$1000 & N \\
SDSS0919+08 & 1.36 & 13,500$\pm$1000 & Y \\
WZ Sge & 1.36 & 14,900$\pm$250 & N \\
SW UMa & 1.36 & 13,900$\pm$900 & N \\
SDSS1035+05 & 1.37 & 10,500$\pm$1000 & N \\
HV Vir & 1.37 & 13,300$\pm$800 & N \\
SDSS1339+48 & 1.38 & 12,500$\pm$1000 & Y \\
SDSS2205+11 & 1.38 & 15,000$\pm$1000 & Y \\
WX Cet & 1.40 & 13,500 & N \\
EG Cnc & 1.41 & 12,300$\pm$700 & N \\
XZ Eri & 1.47 & 15,000$\pm$1500 & N \\
SDSS1514+45 & 1.48? & 10,000$\pm$1000 & Y? \\
VY Aqr & 1.51 & 14,500 & N \\
OY Car & 1.52 & 15,000$\pm$2000 & N \\
SDSS0131-09 & 1.63 & 14,500$\pm$1000 & Y \\
HT Cas & 1.77 & 14,000$\pm$1000 & N \\
REJ1255+26 & 1.99 & 12,000$\pm$1000 & Y \\
EF Peg & 2.00 & 16,600$\pm$1000 & N \\
\enddata
\end{deluxetable}

\clearpage

\begin{figure} [p]
\figurenum {1}
\plotone{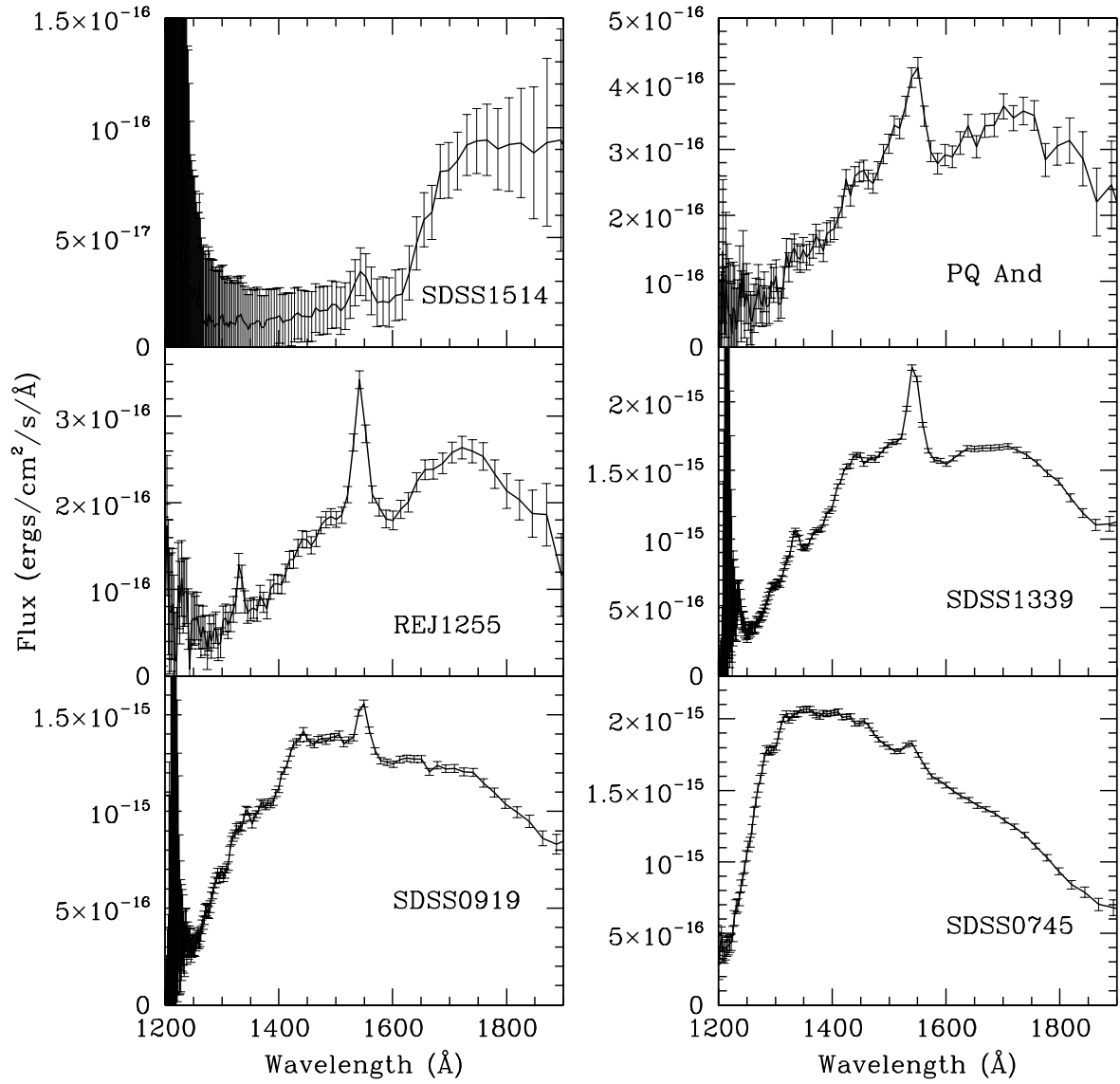}
\caption{HST SBC average spectra of all 6 of our objects using a 17 pixel
extraction.}
\end{figure}

\begin{figure} [p]
\figurenum {2}
\epsscale{0.5}
\plotone{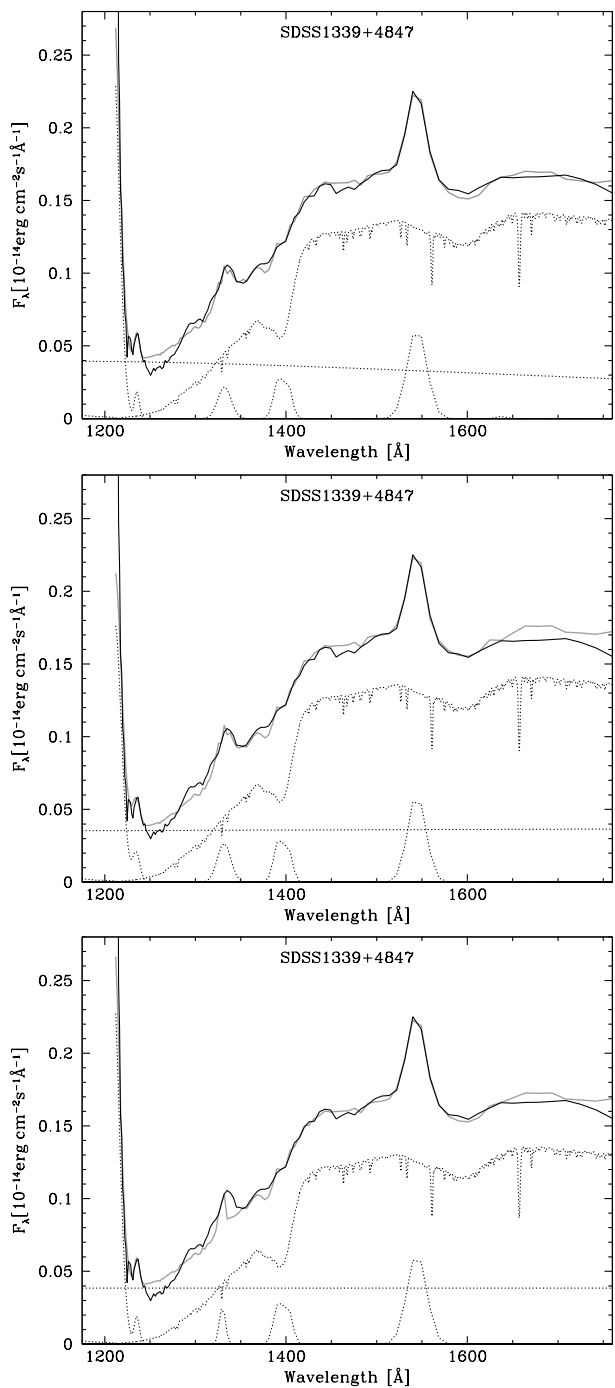}
\caption{White dwarf model (mid curve with absorption lines) fit to the SBC data 
 (dark curve) on SDSS1339+48 with the
disk contribution component (lowest lines) of a black body (top), a power law (mid) and a 
constant
(bottom). In each case, the emission lines are approximated by broad Gaussians
and the sum of the WD, disk and emission lines is the dotted curve on top of the
observed values.}
\end{figure}

\clearpage

\begin{figure} [p]
\figurenum {3}
\epsscale{0.8}
\plotone{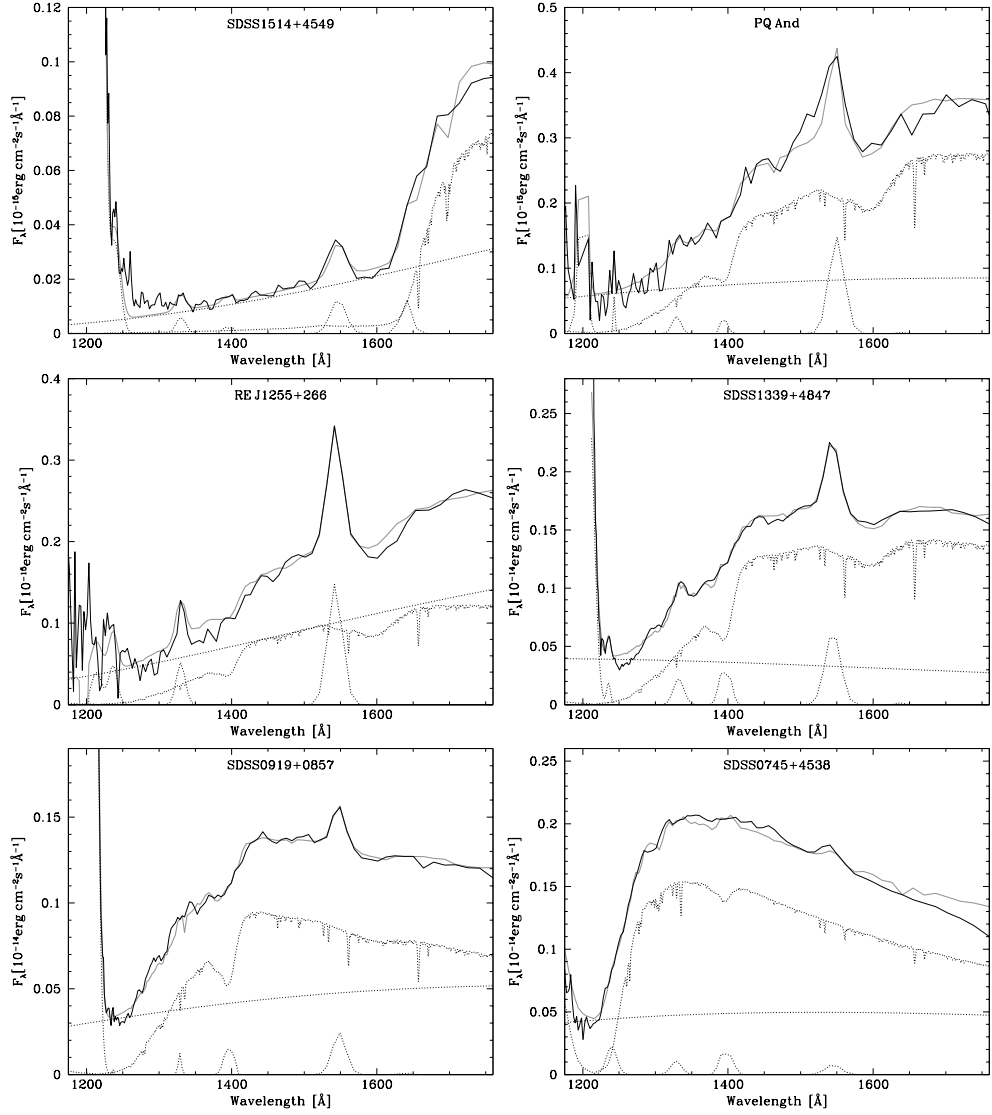}
\caption{White dwarf plus black-body + Gaussian lines model fits to the SBC
data for our 6 objects.}
\end{figure}

\clearpage

\begin{figure} [p]
\figurenum {4}
\epsscale{0.9}
\plotone{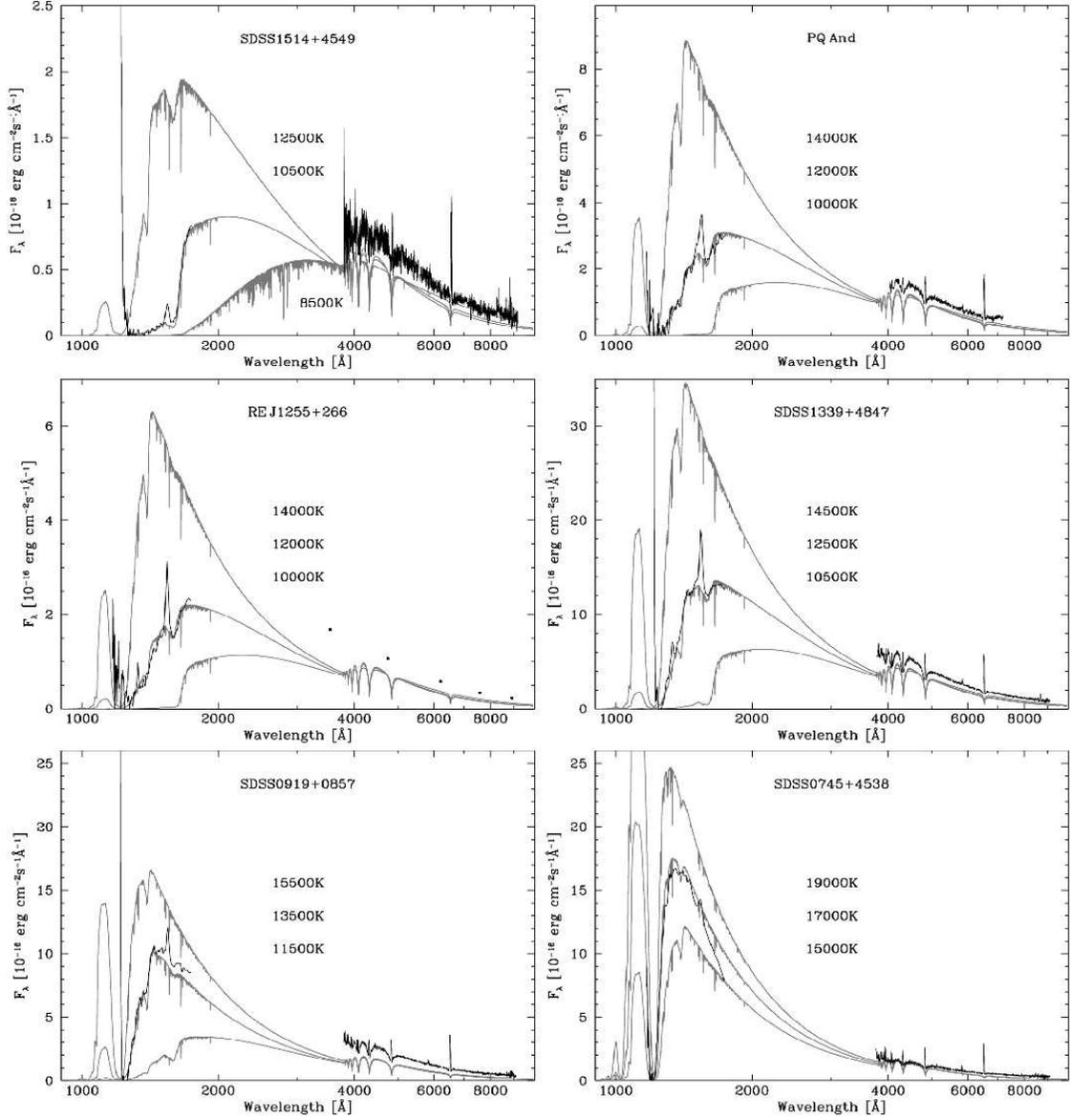}
\caption{SBC data (with constant disk component removed) together with SDSS spectra (ugriz fluxes for REJ1255+26) for our 6 objects shown with dark lines. Grey lines are white dwarf models within 2000K of our best fit temperature white
dwarfs. The difference between the WD model and the observed SDSS data is the disk contribution to the optical. Since SDSS0745+45 had an outburst since
the SDSS spectra were obtained, the optical spectrum may not be representative of true quiescence. }
\end{figure}

\begin{figure} [p]
\figurenum {5}
\includegraphics[width=17cm]{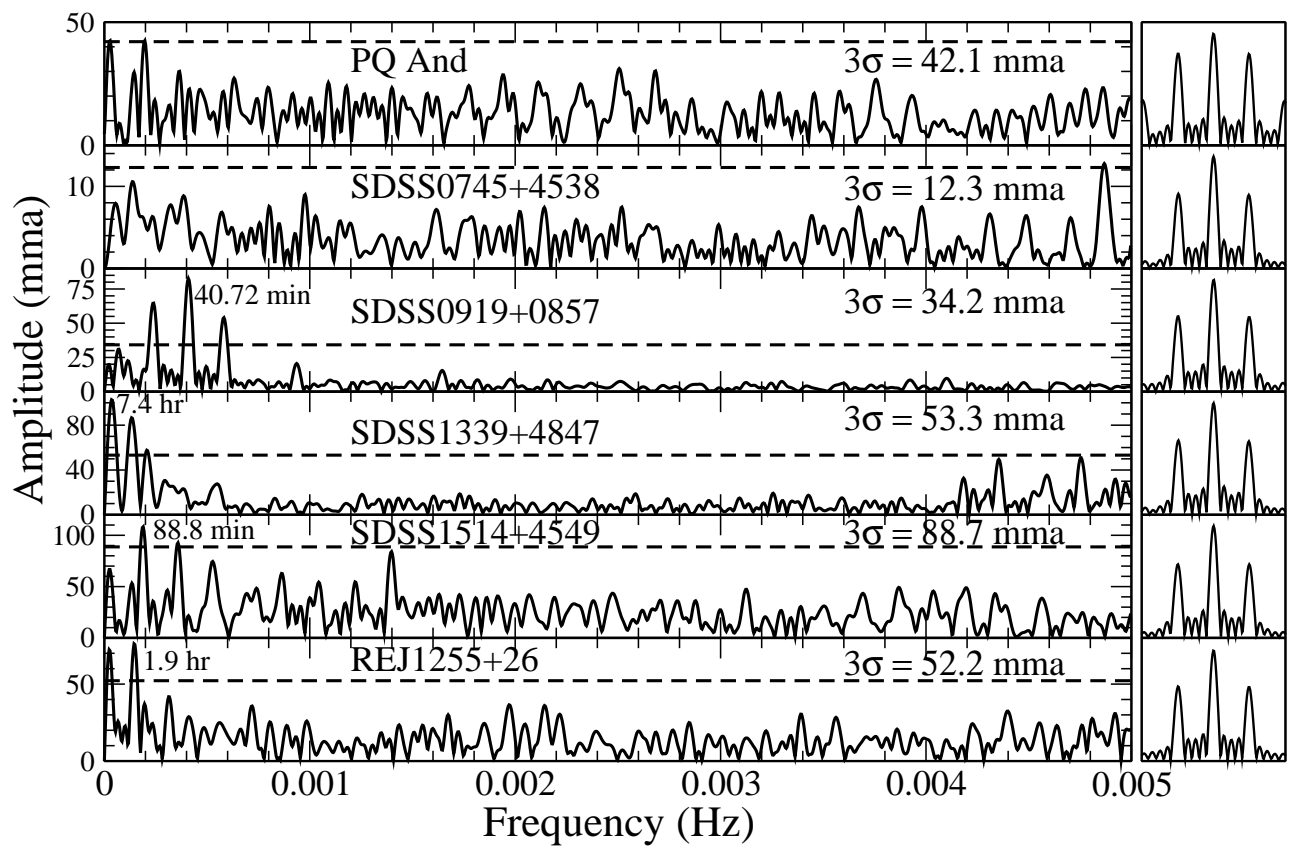}
\caption{DFTs of the HST data for our 6 objects along with a dashed line 
showing the 3$\sigma$ values of the noise for each object. The window functions
are shown in the right-hand column, with the same scaling as the DFT along
the x-axis.}
\end{figure}

\clearpage
\begin{figure} [p]
\figurenum {6}
\epsscale{1.0}
\plotone{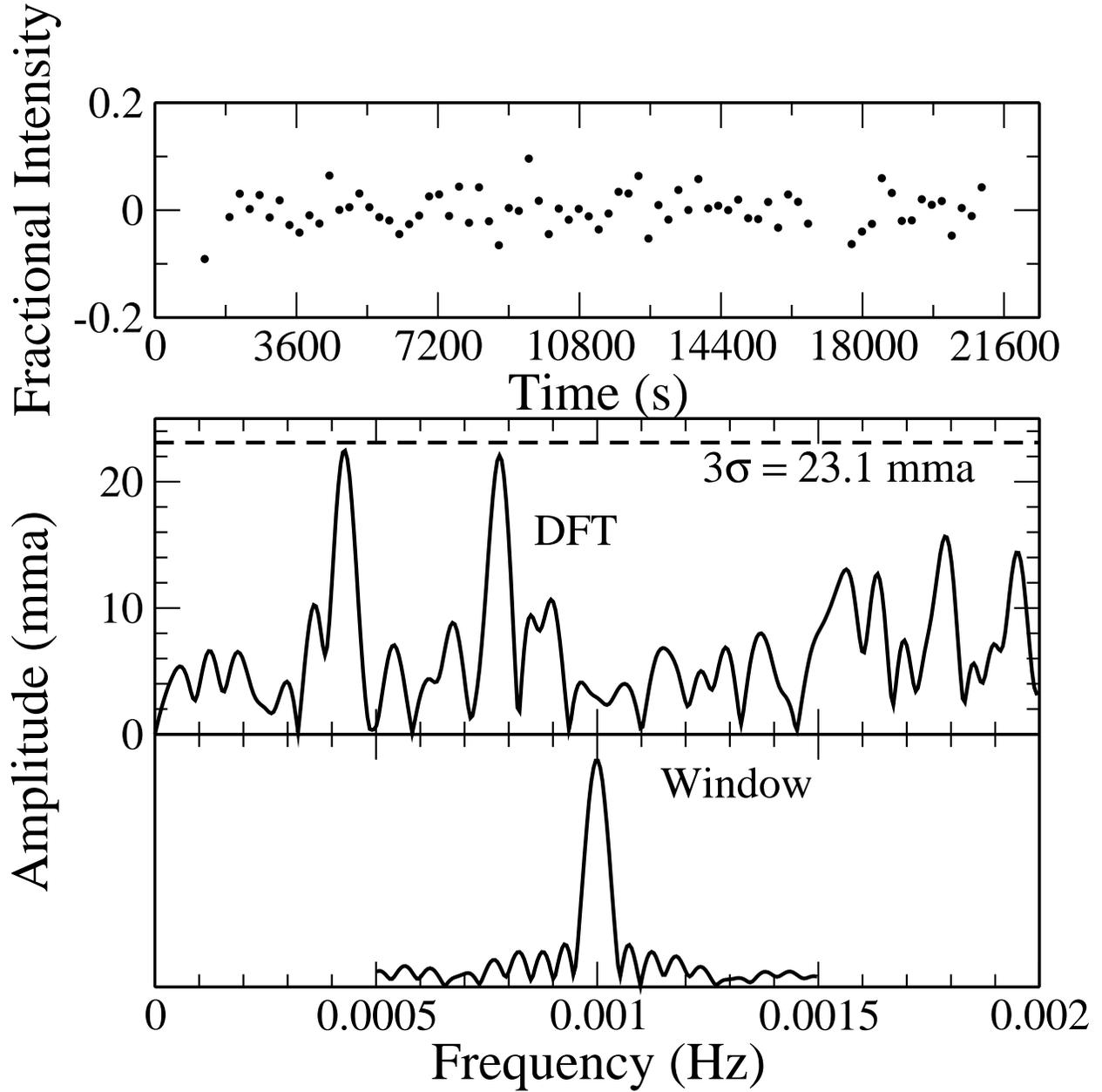}
\caption{Normalized intensity (top), DFT (middle) and window function (bottom)
of the NOFS data on PQ And taken simultaneously with the SBC on 2007 September 13. The two major peaks occur at 2337s and 1285s.} 
\end{figure}
 
\clearpage
\begin{figure} [p]
\figurenum {7}
\epsscale{1.0}
\plotone{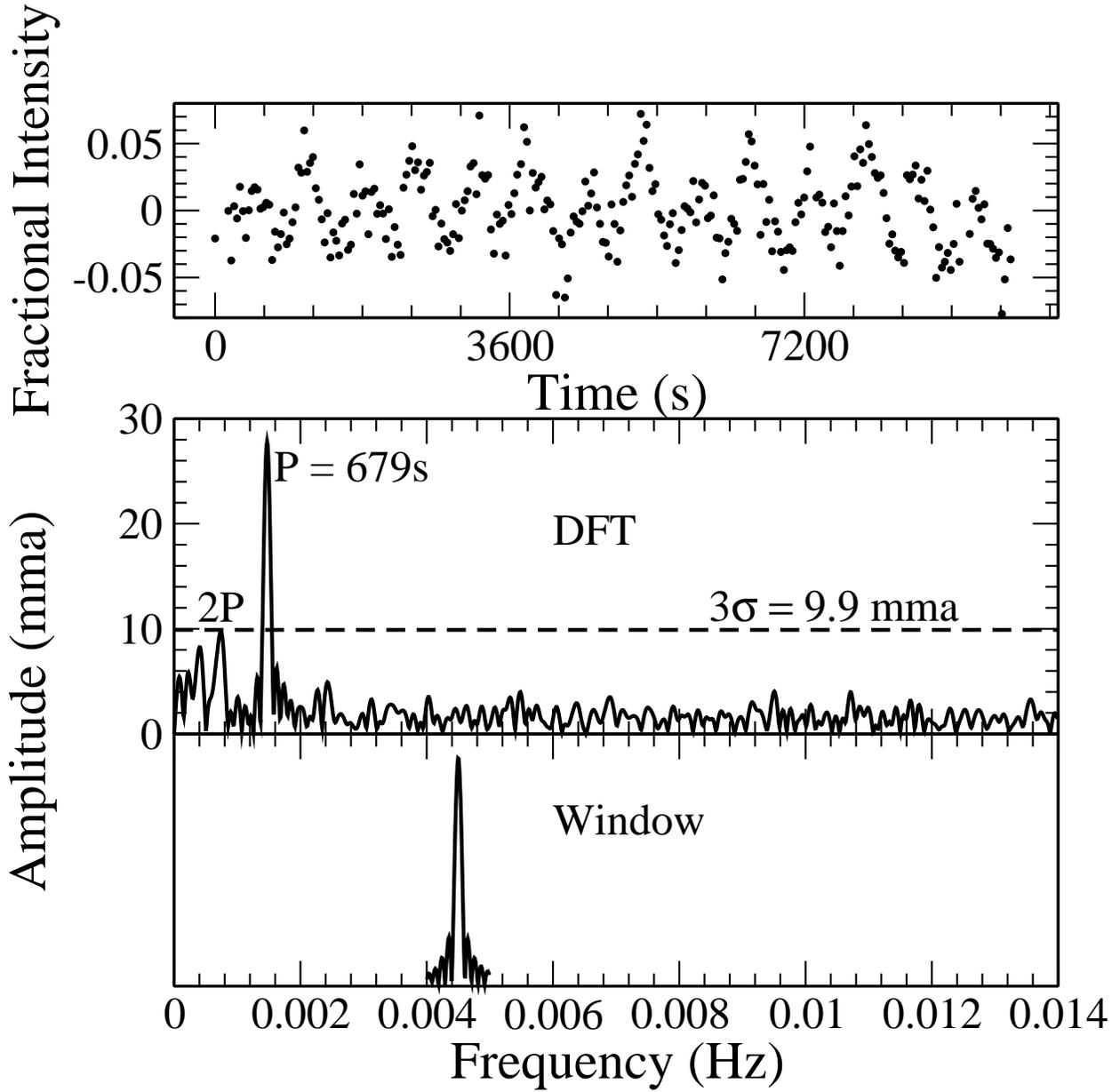}
\caption{Intensity, DFT and window function for the WIYN data on PQ And obtained
one year after the SBC data.}
\end{figure}

\clearpage
\begin{figure} [p]
\figurenum {8}
\epsscale{1.0}
\plotone{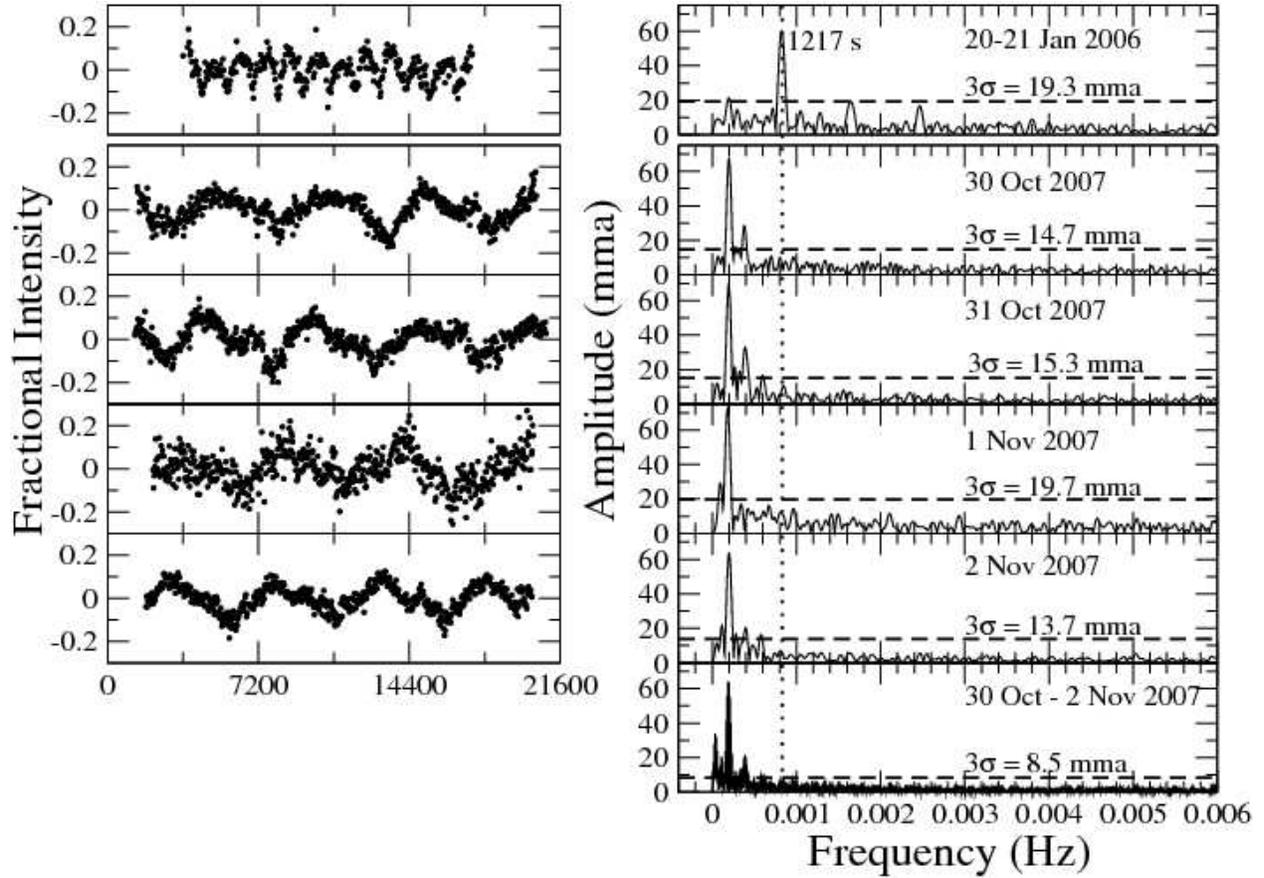}
\caption{Comparison of  SDSS0745+45 data from 2006 with the 4 nights surrounding the 2007 HST observations. Note the visible difference of the light curve as well as the changes in the period evident in the DFTs. Dashed lines show the 3$\sigma$ values for the noise for each dataset.}
\end{figure}

\clearpage
\begin{figure} [p]
\figurenum {9}
\epsscale{1.0}
\plotone{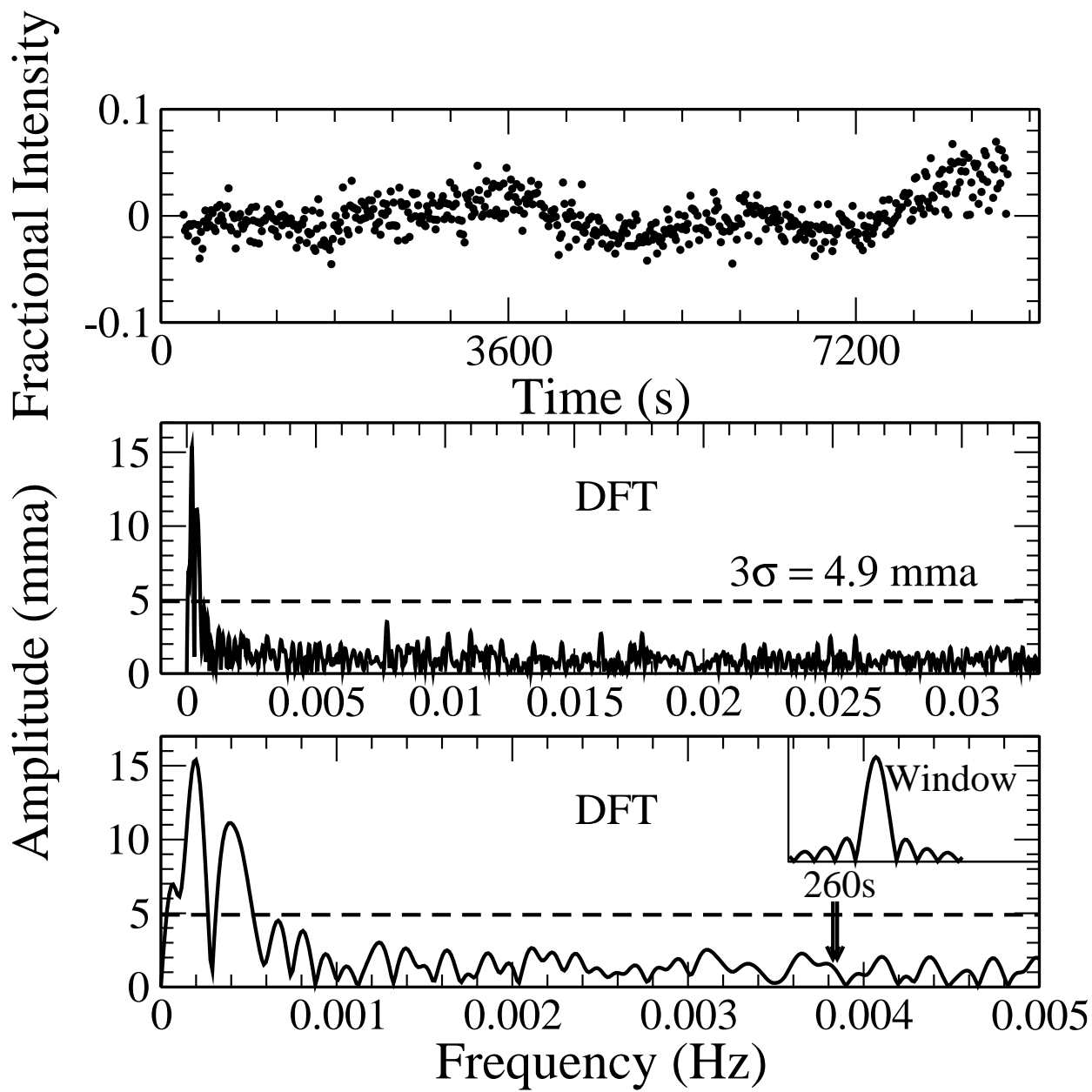}
\caption{Intensity, DFT with window function, and expanded view of low frequencies for McDonald 
Observatory data on SDSS0919+08 obtained two nights prior to the HST data. The position of the
260s pulsation period that was previously observed is marked.} 
\end{figure}

\clearpage
\begin{figure} [p]
\figurenum {10}
\plotone{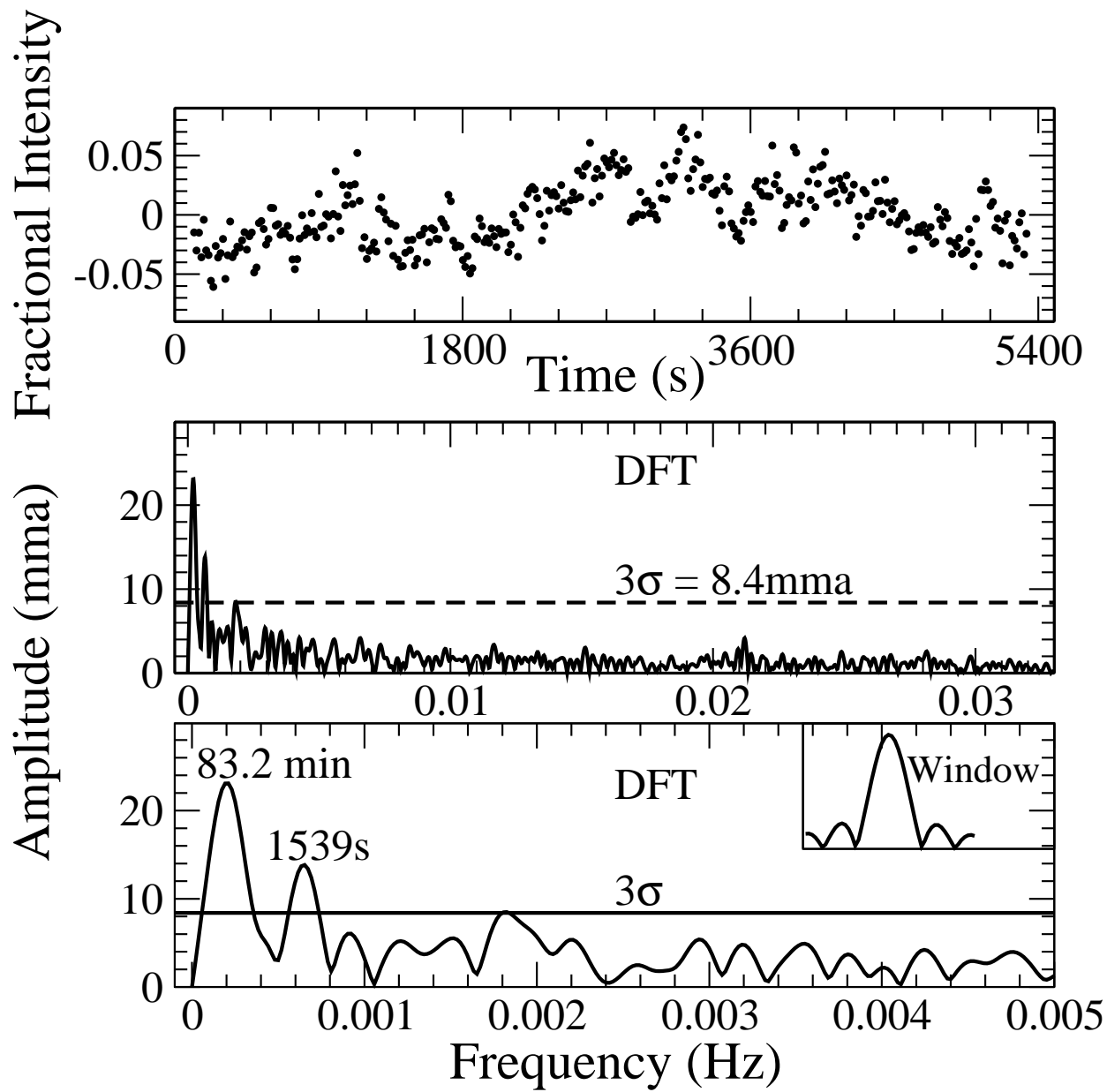}
\caption{Intensity, DFT with noise level marked, and expanded view with 
window function for APO
data on SDSS1339+48 obtained two nights prior to the HST data. The 
previously observed pulsation period at 641s (0.00156 Hz) is not evident.}
\end{figure}

\clearpage
\begin{figure} [p]
\figurenum {11}
\plotone{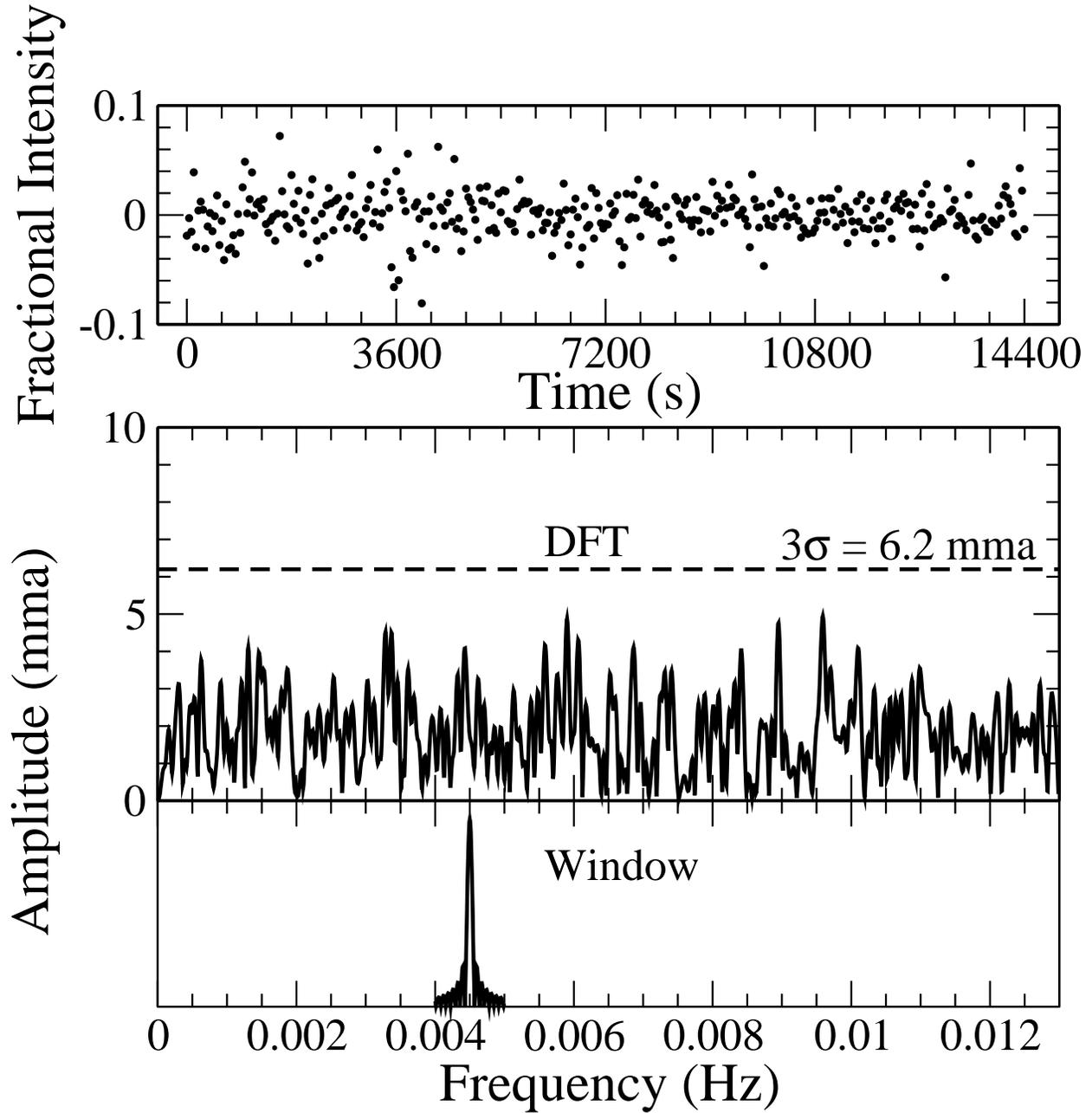}
\caption{Intensity, DFT and window function for APO
data on SDSS1514+45 that overlaps 1.5 hrs with HST observations.
No periods are evident.} 
\end{figure}

\clearpage
\begin{figure} [p]
\figurenum {12}
\plotone{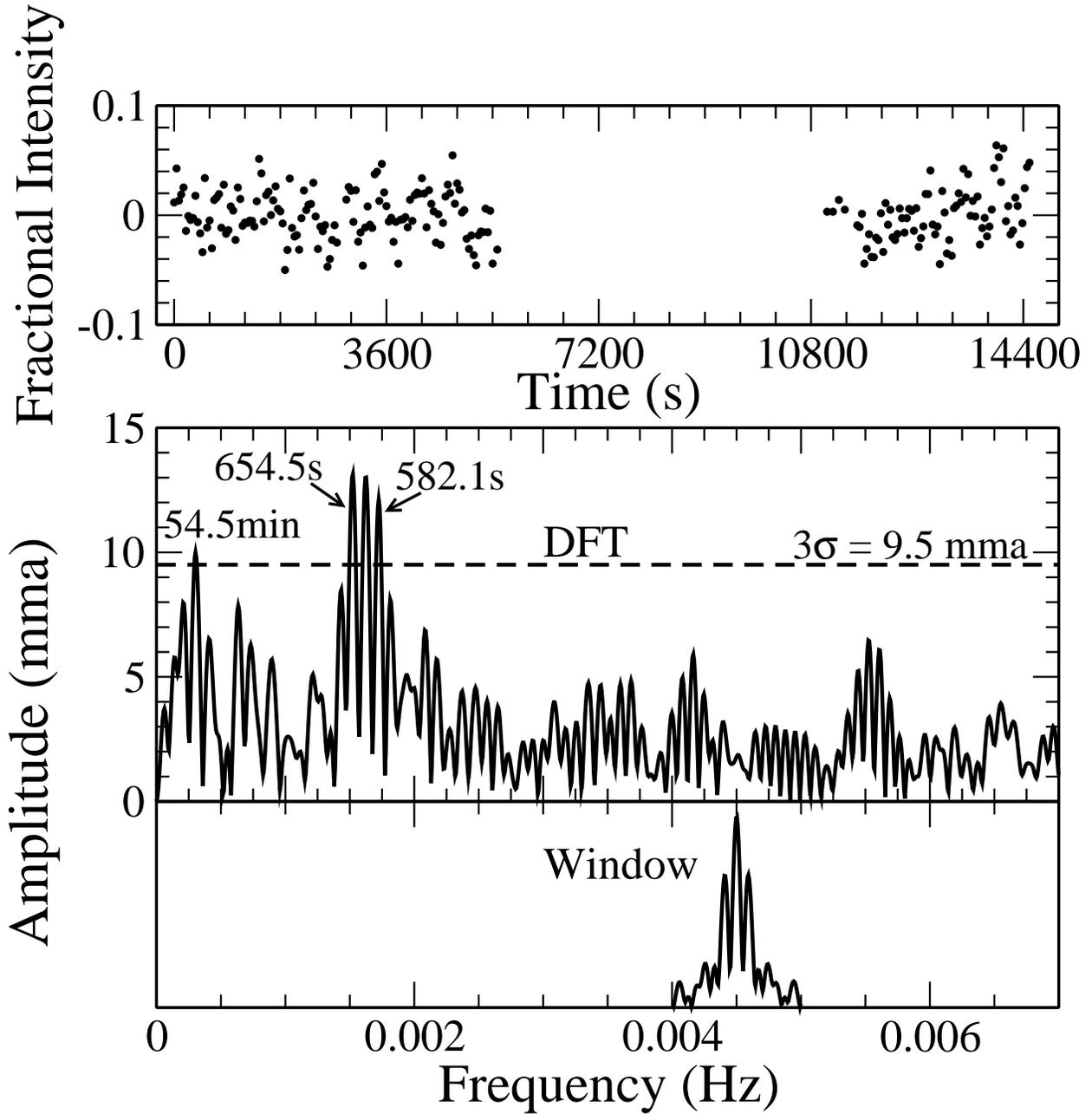}
\caption{Intensity, DFT and window function for APO
data on REJ1255+26 obtained one night after the HST data. 
Two significant short periods are evident.}
\end{figure}

\clearpage
\begin{figure} [p]
\figurenum {13}
\plotone{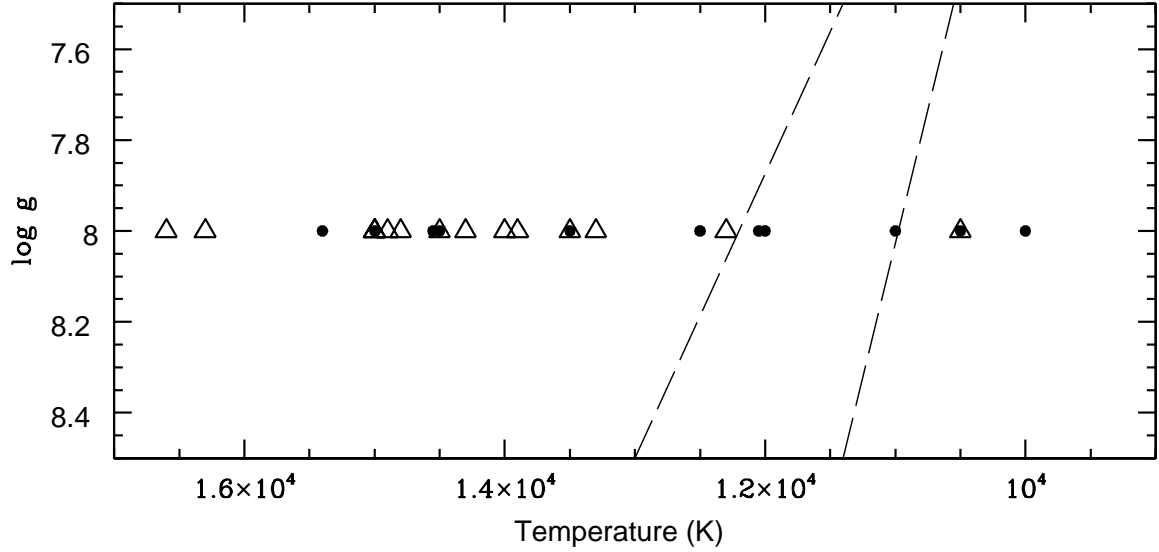}
\caption{Temperatures of accreting pulsators (solid dots) and of non-pulsating
CV white dwarfs (triangles) as a function of log g. The ZZ Ceti instability strip limits (Gianninas et
al. (2007) are shown as dashed lines. The 3 pulsators within the strip are
SDSS1507+52, REJ1255+26 and PQ And. The two objects just outside the blue
edge are the non-pulsator EG Cnc and the pulsator SDSS1339+48.}
\end{figure}

\clearpage
\begin{figure} [p]
\figurenum {14}
\plotone{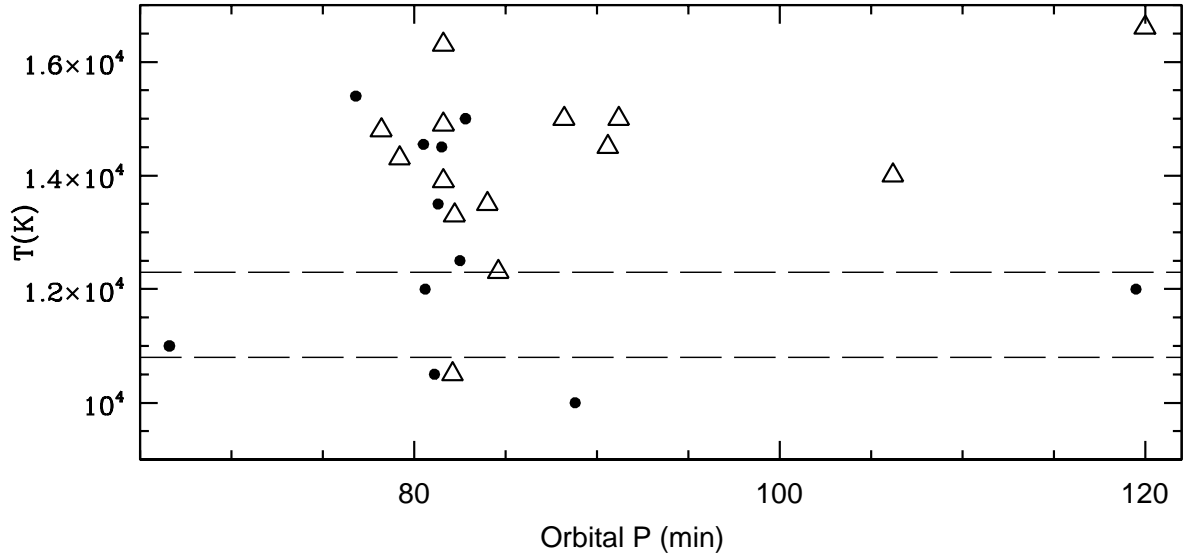}
\caption{Temperatures of accreting pulsators (solid dots) and of non-pulsating
CV white dwarfs (triangles) as a function of orbital period. The limits
of the ZZ Ceti instability strip for log g=8 (Gianninas et al. 2007) are
shown as dashed lines.}
\end{figure}

\end{document}